%\input harvmac

%\input fontch.tex

%%%%%%%%%%%%%%%%%%  tex macros for preprints, cm version %%%%%%%%%%%%%%
%         (P. Ginsparg <ginsparg@lanl.gov>, last updated 7/94)
%                if confused, type `b' in response to query
%           hypertex extensions (still provisional), 7/26/94
% 201-925-0069
%---------------------------------------------------------------------%
%\input hyperbasics %comment out this line to restore non-hyper functionality
%
%% site dependent options:
%% \unredoffs and \redoffs define horizontal and vertical offsets
%% respectively for unreduced and reduced modes. \speclscape defines
%% the \special{} call that sets printer to landscape (sideways) mode.
%% from standard set below, leave uncommented as appropriate or redefine
%
%%% next 400dpi
\def\unredoffs{} \def\redoffs{\voffset=-.31truein\hoffset=-.48truein}
\def\speclscape{}
%\def\speclscape{\special{papersize=11in,8.5in}}
%
%%% apple lw
%\def\unredoffs{} \def\redoffs{\voffset=-.31truein\hoffset=-.59truein}
%\def\speclscape{\special{ps: landscape}}
%
%%% qms lasergrafix:
%\def\unredoffs{} \def\redoffs{\voffset=-.4truein\hoffset=.125truein}
%\def\speclscape{\special{qms: landscape}}
%
%%% saclay A4 paper:
%\def\unredoffs{\hoffset-.14truein\voffset-.2truein}
%\def\redoffs{\voffset=-.45truein\hoffset=-.21truein}
%\def\speclscape{\special{landscape}}
%
%---------------------------------------------------------------------%
%
\newbox\leftpage \newdimen\fullhsize \newdimen\hstitle \newdimen\hsbody
\tolerance=1000\hfuzz=2pt
\catcode`\@=11 % This allows us to modify PLAIN macros.
\ifx\hyperdef\UNd@FiNeD\def\hyperdef#1#2#3#4{#4}\def\hyperref#1#2#3#4{#4}\fi
\def\bigans{b }
\def\answ{b }
%\message{ big or little (b/l)? }\read-1 to\answ
%
\ifx\answ\bigans\message{(This will come out unreduced.}
\magnification=1200\unredoffs\baselineskip=16pt plus 2pt minus 1pt
\hsbody=\hsize \hstitle=\hsize %take default values for unreduced format
\else\message{(This will be reduced.} \let\l@r=L
\magnification=1000\baselineskip=16pt plus 2pt minus 1pt \vsize=7truein
\redoffs \hstitle=8truein\hsbody=4.75truein\fullhsize=10truein\hsize=\hsbody
\output={\ifnum\pageno=0 %%% This is the HUTP version
  \shipout\vbox{\speclscape{\hsize\fullhsize\makeheadline}
    \hbox to \fullhsize{\hfill\pagebody\hfill}}\advancepageno
  \else
  \almostshipout{\leftline{\vbox{\pagebody\makefootline}}}\advancepageno
  \fi}
\def\almostshipout#1{\if L\l@r \count1=1 \message{[\the\count0.\the\count1]}
      \global\setbox\leftpage=#1 \global\let\l@r=R
 \else \count1=2
  \shipout\vbox{\speclscape{\hsize\fullhsize\makeheadline}
      \hbox to\fullhsize{\box\leftpage\hfil#1}}  \global\let\l@r=L\fi}
\fi
%---------------------------------------------------------------------
%
\newcount\yearltd\yearltd=\year\advance\yearltd by -2000

\def\Title#1#2{\nopagenumbers\abstractfont\hsize=\hstitle\rightline{#1}%
\vskip 1in\centerline{\titlefont #2}\abstractfont\vskip .5in\pageno=0}
\def\Date#1{\vfill\leftline{#1}\tenpoint\supereject\global\hsize=\hsbody%
\footline={\hss\tenrm\hyperdef\hypernoname{page}\folio\folio\hss}}%
% (restores pagenumbers)
%
%       use following instead of \Date on the preliminary draft,
%       puts date/time on each page in big mode, writes labels in margins

\def\draftmode{\message{ DRAFTMODE }\def\draftdate{{\rm preliminary draft:
\number\month/\number\day/\number\yearltd\ \ \hourmin}}%
\headline={\hfil\draftdate}\writelabels\baselineskip=20pt plus 2pt minus 2pt
 {\count255=\time\divide\count255 by 60 \xdef\hourmin{\number\count255}
  \multiply\count255 by-60\advance\count255 by\time
  \xdef\hourmin{\hourmin:\ifnum\count255<10 0\fi\the\count255}}}
%       use \nolabels to get rid of eqn, ref, and fig labels in draft mode
\def\nolabels{\def\wrlabeL##1{}\def\eqlabeL##1{}\def\reflabeL##1{}}
\def\writelabels{\def\wrlabeL##1{\leavevmode\vadjust{\rlap{\smash%
{\line{{\escapechar=` \hfill\rlap{\sevenrm\hskip.03in\string##1}}}}}}}%
\def\eqlabeL##1{{\escapechar-1\rlap{\sevenrm\hskip.05in\string##1}}}%
\def\reflabeL##1{\noexpand\llap{\noexpand\sevenrm\string\string\string##1}}}
\nolabels
%
% tagged sec numbers
\global\newcount\secno \global\secno=0
\global\newcount\meqno \global\meqno=1
\def\s@csym{}
\def\newsec#1{\global\advance\secno by1%
{\toks0{#1}\message{(\the\secno. \the\toks0)}}%
%\ifx\answ\bigans \vfill\eject \else \bigbreak\bigskip \fi  %if desired
\global\subsecno=0\eqnres@t\let\s@csym\secsym\xdef\secn@m{\the\secno}\noindent
{\bf\hyperdef\hypernoname{section}{\the\secno}{\the\secno.} #1}%
\writetoca{{\string\hyperref{}{section}{\the\secno}{\the\secno.}} {#1}}%
\par\nobreak\medskip\nobreak}
\def\eqnres@t{\xdef\secsym{\the\secno.}\global\meqno=1\bigbreak\bigskip}
\def\sequentialequations{\def\eqnres@t{\bigbreak}}\xdef\secsym{}
\global\newcount\subsecno \global\subsecno=0
\def\subsec#1{\global\advance\subsecno by1%
{\toks0{#1}\message{(\s@csym\the\subsecno. \the\toks0)}}%
\ifnum\lastpenalty>9000\else\bigbreak\fi
\noindent{\it\hyperdef\hypernoname{subsection}{\secn@m.\the\subsecno}%
{\secn@m.\the\subsecno.} #1}\writetoca{\string\quad
{\string\hyperref{}{subsection}{\secn@m.\the\subsecno}{\secn@m.\the\subsecno.}}
{#1}}\par\nobreak\medskip\nobreak}
\def\appendix#1#2{\global\meqno=1\global\subsecno=0\xdef\secsym{\hbox{#1.}}%
\bigbreak\bigskip\noindent{\bf Appendix \hyperdef\hypernoname{appendix}{#1}%
{#1.} #2}{\toks0{(#1. #2)}\message{\the\toks0}}%
\xdef\s@csym{#1.}\xdef\secn@m{#1}%
\writetoca{\string\hyperref{}{appendix}{#1}{Appendix {#1.}} {#2}}%
\par\nobreak\medskip\nobreak}
%
%       \eqn\label{a+b=c}	gives displayed equation, numbered
%				consecutively within sections.
%     \eqnn and \eqna define labels in advance (of eqalign?)
%
\def\checkm@de#1#2{\ifmmode{\def\f@rst##1{##1}\hyperdef\hypernoname{equation}%
{#1}{#2}}\else\hyperref{}{equation}{#1}{#2}\fi}
\def\eqnn#1{\DefWarn#1\xdef #1{(\noexpand\relax\noexpand\checkm@de%
{\s@csym\the\meqno}{\secsym\the\meqno})}%
\wrlabeL#1\writedef{#1\leftbracket#1}\global\advance\meqno by1}
\def\f@rst#1{\c@t#1a\em@ark}\def\c@t#1#2\em@ark{#1}
\def\eqna#1{\DefWarn#1\wrlabeL{#1$\{\}$}%
\xdef #1##1{(\noexpand\relax\noexpand\checkm@de%
{\s@csym\the\meqno\noexpand\f@rst{##1}}{\hbox{$\secsym\the\meqno##1$}})}
\writedef{#1\numbersign1\leftbracket#1{\numbersign1}}\global\advance\meqno by1}
\def\eqn#1#2{\DefWarn#1%
\xdef #1{(\noexpand\hyperref{}{equation}{\s@csym\the\meqno}%
{\secsym\the\meqno})}$$#2\eqno(\hyperdef\hypernoname{equation}%
{\s@csym\the\meqno}{\secsym\the\meqno})\eqlabeL#1$$%
\writedef{#1\leftbracket#1}\global\advance\meqno by1}
\def\xeqn{\expandafter\xe@n}\def\xe@n(#1){#1}
\def\xeqna#1{\expandafter\xe@n#1}
\def\eqns#1{(\e@ns #1{\hbox{}})}
\def\e@ns#1{\ifx\UNd@FiNeD#1\message{eqnlabel \string#1 is undefined.}%
\xdef#1{(?.?)}\fi{\let\hyperref=\relax\xdef\next{#1}}%
\ifx\next\em@rk\def\next{}\else%
\ifx\next#1\xeqn#1\else\def\n@xt{#1}\ifx\n@xt\next#1\else\xeqna#1\fi
\fi\let\next=\e@ns\fi\next}

\def\DefWarn#1{\ifx\UNd@FiNeD#1\else
\immediate\write16{*** WARNING: the label \string#1 is already defined ***}\fi}
%
%			 footnotes
\newskip\footskip\footskip14pt plus 1pt minus 1pt %sets footnote baselineskip
\def\footnotefont{\ninepoint}\def\f@t#1{\footnotefont #1\@foot}
\def\f@@t{\baselineskip\footskip\bgroup\footnotefont\aftergroup\@foot\let\next}
\setbox\strutbox=\hbox{\vrule height9.5pt depth4.5pt width0pt}
\global\newcount\ftno \global\ftno=0
\def\foot{\global\advance\ftno by1\def\foot@rg{\hyperref{}{footnote}%
{\the\ftno}{\the\ftno}\xdef\foot@rg{\noexpand\hyperdef\noexpand\hypernoname%
{footnote}{\the\ftno}{\the\ftno}}}\footnote{$^{\foot@rg}$}}
%
%say \footend to put footnotes at end
%will cause problems if \ref used inside \foot, instead use \nref before
\newwrite\ftfile
\def\footend{\def\foot{\global\advance\ftno by1\chardef\wfile=\ftfile
%%$^{\the\ftno}$\ifnum\ftno=1\immediate\openout\ftfile=\jobname.fts\fi%
\hyperref{}{footnote}{\the\ftno}{$^{\the\ftno}$}%
\ifnum\ftno=1\immediate\openout\ftfile=\jobname.fts\fi%
\immediate\write\ftfile{\noexpand\smallskip%
%%\noexpand\item{f\the\ftno:\ }\pctsign}\findarg}%
\noexpand\item{\noexpand\hyperdef\noexpand\hypernoname{footnote}
{\the\ftno}{f\the\ftno}:\ }\pctsign}\findarg}%
\def\footatend{\vfill\eject\immediate\closeout\ftfile{\parindent=20pt
\centerline{\bf Footnotes}\nobreak\bigskip\input \jobname.fts }}}
\def\footatend{}
%
%     \ref\label{text}
% generates a number, assigns it to \label, generates an entry.
% To list the refs on a separate page,  \listrefs
%
\global\newcount\refno \global\refno=1
\newwrite\rfile
\def\ref{[\hyperref{}{reference}{\the\refno}{\the\refno}]\nref}
\def\nref#1{\DefWarn#1%
\xdef#1{[\noexpand\hyperref{}{reference}{\the\refno}{\the\refno}]}%
\writedef{#1\leftbracket#1}%
\ifnum\refno=1\immediate\openout\rfile=\jobname.refs\fi
\chardef\wfile=\rfile\immediate\write\rfile{\noexpand\item{[\noexpand\hyperdef%
\noexpand\hypernoname{reference}{\the\refno}{\the\refno}]\ }%
\reflabeL{#1\hskip.31in}\pctsign}\global\advance\refno by1\findarg}
%	horrible hack to sidestep tex \write limitation
\def\findarg#1#{\begingroup\obeylines\newlinechar=`\^^M\pass@rg}
{\obeylines\gdef\pass@rg#1{\writ@line\relax #1^^M\hbox{}^^M}%
\gdef\writ@line#1^^M{\expandafter\toks0\expandafter{\striprel@x #1}%
\edef\next{\the\toks0}\ifx\next\em@rk\let\next=\endgroup\else\ifx\next\empty%
\else\immediate\write\wfile{\the\toks0}\fi\let\next=\writ@line\fi\next\relax}}
\def\striprel@x#1{} \def\em@rk{\hbox{}}
\def\lref{\begingroup\obeylines\lr@f}
\def\lr@f#1#2{\DefWarn#1\gdef#1{\let#1=\UNd@FiNeD\ref#1{#2}}\endgroup\unskip}

\def\addref#1{\immediate\write\rfile{\noexpand\item{}#1}} %now unnecessary
\def\listrefs{\footatend\vfill\supereject\immediate\closeout\rfile\writestoppt
\baselineskip=\footskip\centerline{{\bf References}}\bigskip{\parindent=20pt%
\frenchspacing\escapechar=` \input \jobname.refs\vfill\eject}\nonfrenchspacing}
\def\startrefs#1{\immediate\openout\rfile=\jobname.refs\refno=#1}
\def\xref{\expandafter\xr@f}\def\xr@f[#1]{#1}
\def\refs#1{\count255=1[\r@fs #1{\hbox{}}]}
\def\r@fs#1{\ifx\UNd@FiNeD#1\message{reflabel \string#1 is undefined.}%
\nref#1{need to supply reference \string#1.}\fi%
\vphantom{\hphantom{#1}}{\let\hyperref=\relax\xdef\next{#1}}%
\ifx\next\em@rk\def\next{}%
\else\ifx\next#1\ifodd\count255\relax\xref#1\count255=0\fi%
\else#1\count255=1\fi\let\next=\r@fs\fi\next}
%

%
% this is ugly, but moore insists
\newwrite\ffile\global\newcount\figno \global\figno=1
\def\fig{fig.~\hyperref{}{figure}{\the\figno}{\the\figno}\nfig}
\def\nfig#1{\DefWarn#1%
\xdef#1{fig.~\noexpand\hyperref{}{figure}{\the\figno}{\the\figno}}%
\writedef{#1\leftbracket fig.\noexpand~\xfig#1}%
\ifnum\figno=1\immediate\openout\ffile=\jobname.figs\fi\chardef\wfile=\ffile%
{\let\hyperref=\relax
\immediate\write\ffile{\noexpand\medskip\noexpand\item{Fig.\ %
\noexpand\hyperdef\noexpand\hypernoname{figure}{\the\figno}{\the\figno}. }
\reflabeL{#1\hskip.55in}\pctsign}}\global\advance\figno by1\findarg}
\def\listfigs{\vfill\eject\immediate\closeout\ffile{\parindent40pt
\baselineskip14pt\centerline{{\bf Figure Captions}}\nobreak\medskip
\escapechar=` \input \jobname.figs\vfill\eject}}
\def\xfig{\expandafter\xf@g}\def\xf@g fig.\penalty\@M\ {}
\def\figs#1{figs.~\f@gs #1{\hbox{}}}
\def\f@gs#1{{\let\hyperref=\relax\xdef\next{#1}}\ifx\next\em@rk\def\next{}\else
\ifx\next#1\xfig #1\else#1\fi\let\next=\f@gs\fi\next}
\def\figin{\epsfcheck\figin}\def\figins{\epsfcheck\figins}
\def\epsfcheck{\ifx\epsfbox\UNd@FiNeD
\message{(NO epsf.tex, FIGURES WILL BE IGNORED)}
\gdef\figin##1{\vskip2in}\gdef\figins##1{\hskip.5in}% blank space instead
\else\message{(FIGURES WILL BE INCLUDED)}%
\gdef\figin##1{##1}\gdef\figins##1{##1}\fi}
\def\DefWarn#1{}
\def\figinsert{\goodbreak\midinsert}
\def\ifig#1#2#3{\DefWarn#1\xdef#1{fig.~\noexpand\hyperref{}{figure}%
{\the\figno}{\the\figno}}\writedef{#1\leftbracket fig.\noexpand~\xfig#1}%
\figinsert\figin{\centerline{#3}}\medskip\centerline{\vbox{\baselineskip12pt
\advance\hsize by -1truein\noindent\wrlabeL{#1=#1}\footnotefont%
{\bf Fig.~\hyperdef\hypernoname{figure}{\the\figno}{\the\figno}:} #2}}
\bigskip\endinsert\global\advance\figno by1}
\newwrite\lfile
{\escapechar-1\xdef\pctsign{\string\%}\xdef\leftbracket{\string\{}
\xdef\rightbracket{\string\}}\xdef\numbersign{\string\#}}
\def\writedefs{\immediate\openout\lfile=\jobname.defs \def\writedef##1{%
{\let\hyperref=\relax\let\hyperdef=\relax\let\hypernoname=\relax
 \immediate\write\lfile{\string\def\string##1\rightbracket}}}}%
\def\writestop{\def\writestoppt{\immediate\write\lfile{\string\pageno
 \the\pageno\string\startrefs\leftbracket\the\refno\rightbracket
 \string\def\string\secsym\leftbracket\secsym\rightbracket
 \string\secno\the\secno\string\meqno\the\meqno}\immediate\closeout\lfile}}
\def\writestoppt{}\def\writedef#1{}
\def\seclab#1{\DefWarn#1%
\xdef #1{\noexpand\hyperref{}{section}{\the\secno}{\the\secno}}%
\writedef{#1\leftbracket#1}\wrlabeL{#1=#1}}
\def\subseclab#1{\DefWarn#1%
\xdef #1{\noexpand\hyperref{}{subsection}{\secn@m.\the\subsecno}%
{\secn@m.\the\subsecno}}\writedef{#1\leftbracket#1}\wrlabeL{#1=#1}}
\def\applab#1{\DefWarn#1%
\xdef #1{\noexpand\hyperref{}{appendix}{\secn@m}{\secn@m}}%
\writedef{#1\leftbracket#1}\wrlabeL{#1=#1}}
\newwrite\tfile \def\writetoca#1{}
\def\leaderfill{\leaders\hbox to 1em{\hss.\hss}\hfill}
%	use this to write file with table of contents
\def\writetoc{\immediate\openout\tfile=\jobname.toc
   \def\writetoca##1{{\edef\next{\write\tfile{\noindent ##1
   \string\leaderfill {\string\hyperref{}{page}{\noexpand\number\pageno}%
                       {\noexpand\number\pageno}} \par}}\next}}}
%       and this lists table of contents on second pass
\newread\ch@ckfile
\def\listtoc{\immediate\closeout\tfile\immediate\openin\ch@ckfile=\jobname.toc
\ifeof\ch@ckfile\message{no file \jobname.toc, no table of contents this pass}%
\else\closein\ch@ckfile\centerline{\bf Contents}\nobreak\medskip%
{\baselineskip=12pt\footnotefont\parskip=0pt\catcode`\@=11\input\jobname.toc
\catcode`\@=12\bigbreak\bigskip}\fi}
\catcode`\@=12 % at signs are no longer letters
%
%	Unpleasantness in calling in abstract and title fonts
\edef\tfontsize{\ifx\answ\bigans scaled\magstep3\else scaled\magstep4\fi}
\font\titlerm=cmr10 \tfontsize \font\titlerms=cmr7 \tfontsize
\font\titlermss=cmr5 \tfontsize \font\titlei=cmmi10 \tfontsize
\font\titleis=cmmi7 \tfontsize \font\titleiss=cmmi5 \tfontsize
\font\titlesy=cmsy10 \tfontsize \font\titlesys=cmsy7 \tfontsize
\font\titlesyss=cmsy5 \tfontsize \font\titleit=cmti10 \tfontsize
\skewchar\titlei='177 \skewchar\titleis='177 \skewchar\titleiss='177
\skewchar\titlesy='60 \skewchar\titlesys='60 \skewchar\titlesyss='60
\def\titlefont{\def\rm{\fam0\titlerm}% switch to title font
\textfont0=\titlerm \scriptfont0=\titlerms \scriptscriptfont0=\titlermss
\textfont1=\titlei \scriptfont1=\titleis \scriptscriptfont1=\titleiss
\textfont2=\titlesy \scriptfont2=\titlesys \scriptscriptfont2=\titlesyss
\textfont\itfam=\titleit \def\it{\fam\itfam\titleit}\rm}
 \ifx\answ\bigans\else scaled\magstep1\fi
\ifx\answ\bigans\def\abstractfont{\tenpoint}\else
\font\absit=cmti10 scaled \magstep1
\font\abssl=cmsl10 scaled \magstep1
\font\absrm=cmr10 scaled\magstep1 \font\absrms=cmr7 scaled\magstep1
\font\absrmss=cmr5 scaled\magstep1 \font\absi=cmmi10 scaled\magstep1
\font\absis=cmmi7 scaled\magstep1 \font\absiss=cmmi5 scaled\magstep1
\font\abssy=cmsy10 scaled\magstep1 \font\abssys=cmsy7 scaled\magstep1
\font\abssyss=cmsy5 scaled\magstep1 \font\absbf=cmbx10 scaled\magstep1
\skewchar\absi='177 \skewchar\absis='177 \skewchar\absiss='177
\skewchar\abssy='60 \skewchar\abssys='60 \skewchar\abssyss='60
\def\abstractfont{\def\rm{\fam0\absrm}% switch to abstract font
\textfont0=\absrm \scriptfont0=\absrms \scriptscriptfont0=\absrmss
\textfont1=\absi \scriptfont1=\absis \scriptscriptfont1=\absiss
\textfont2=\abssy \scriptfont2=\abssys \scriptscriptfont2=\abssyss
\textfont\itfam=\absit \def\it{\fam\itfam\absit}\def\footnotefont{\tenpoint}%
\textfont\slfam=\abssl \def\sl{\fam\slfam\abssl}%
\textfont\bffam=\absbf \def\bf{\fam\bffam\absbf}\rm}\fi
\def\tenpoint{\def\rm{\fam0\tenrm}% switch back to 10-point type
\textfont0=\tenrm \scriptfont0=\sevenrm \scriptscriptfont0=\fiverm
\textfont1=\teni  \scriptfont1=\seveni  \scriptscriptfont1=\fivei
\textfont2=\tensy \scriptfont2=\sevensy \scriptscriptfont2=\fivesy
\textfont\itfam=\tenit \def\it{\fam\itfam\tenit}\def\footnotefont{\ninepoint}%
\textfont\bffam=\tenbf \def\bf{\fam\bffam\tenbf}\def\sl{\fam\slfam\tensl}\rm}
\font\ninerm=cmr9 \font\sixrm=cmr6 \font\ninei=cmmi9 \font\sixi=cmmi6
\font\ninesy=cmsy9 \font\sixsy=cmsy6 \font\ninebf=cmbx9
\font\nineit=cmti9 \font\ninesl=cmsl9 \skewchar\ninei='177
\skewchar\sixi='177 \skewchar\ninesy='60 \skewchar\sixsy='60
\def\ninepoint{\def\rm{\fam0\ninerm}% switch to footnote font
\textfont0=\ninerm \scriptfont0=\sixrm \scriptscriptfont0=\fiverm
\textfont1=\ninei \scriptfont1=\sixi \scriptscriptfont1=\fivei
\textfont2=\ninesy \scriptfont2=\sixsy \scriptscriptfont2=\fivesy
\textfont\itfam=\ninei \def\it{\fam\itfam\nineit}\def\sl{\fam\slfam\ninesl}%
\textfont\bffam=\ninebf \def\bf{\fam\bffam\ninebf}\rm}
%
%---------------------------------------------------------------------
%

\hyphenation{anom-aly anom-alies coun-ter-term coun-ter-terms}
\def\inv{^{\raise.15ex\hbox{${\scriptscriptstyle -}$}\kern-.05em 1}}

\def\Dsl{\,\raise.15ex\hbox{/}\mkern-13.5mu D} %this one can be subscripted
\def\dsl{\raise.15ex\hbox{/}\kern-.57em\partial}

 %pound sterling
\def\lspace{\ifx\answ\bigans{}\else\qquad\fi}
\def\lbspace{\ifx\answ\bigans{}\else\hskip-.2in\fi} % $$\lbspace...$$
\def\boxeqn#1{\vcenter{\vbox{\hrule\hbox{\vrule\kern3pt\vbox{\kern3pt
	\hbox{${\displaystyle #1}$}\kern3pt}\kern3pt\vrule}\hrule}}}
\def\mbox#1#2{\vcenter{\hrule \hbox{\vrule height#2in
		\kern#1in \vrule} \hrule}}  %e.g. \mbox{.1}{.1}
%	matters of taste
%\def\tilde{\widetilde} \def\bar{\overline} \def\hat{\widehat}
%
% some sample definitions
  %     curly letters

\def\darr#1{\raise1.5ex\hbox{$\leftrightarrow$}\mkern-16.5mu #1}
 %pound sterling

 %puts a small half in a displayed eqn
\def\roughly#1{\raise.3ex\hbox{$#1$\kern-.75em\lower1ex\hbox{$\sim$}}}

\def\smallfig#1#2#3{\DefWarn#1\xdef#1{fig.~\the\figno}
\writedef{#1\leftbracket fig.\noexpand~\the\figno}%
\figinsert\figin{\centerline{#3}}\medskip\centerline{\vbox{
\baselineskip12pt\advance\hsize by -1truein
\noindent\footnotefont{\bf Fig.~\the\figno:} #2}}
\endinsert\global\advance\figno by1}

\input amssym

%
% for ordinary tex
%
\ifx\pdfoutput\undefined
\input epsf
\def\fig#1{\epsfbox{#1.eps}}

%
% for pdftex
%
\else
\def\fig#1{\pdfximage {#1.pdf}\pdfrefximage\pdflastximage}

\fi

\def\IZ{\relax\ifmmode\mathchoice
{\hbox{\cmss Z\kern-.4em Z}}{\hbox{\cmss Z\kern-.4em Z}} {\lower.9pt\hbox{\cmsss Z\kern-.4em Z}}
{\lower1.2pt\hbox{\cmsss Z\kern-.4em Z}}\else{\cmss Z\kern-.4em Z}\fi}

\newif\ifdraft\draftfalse
%\drafttrue
\newif\ifinter\interfalse
%\intertrue
\ifdraft\draftmode\else\interfalse\fi
\def\journal#1&#2(#3){\unskip, \sl #1\ \bf #2 \rm(19#3) }
\def\andjournal#1&#2(#3){\sl #1~\bf #2 \rm (19#3) }

\def\frac#1#2{{#1\over#2}}

\def\inbar{\,\vrule height1.5ex width.4pt depth0pt}
\def\IC{\relax\hbox{$\inbar\kern-.3em{\rm C}$}}
\def\IR{\relax{\rm I\kern-.18em R}}
\def\IP{\relax{\rm I\kern-.18em P}}

%
%%%%%%%%%%%%%%%%%%%%%%%%%%%%%%%%%%%%
%

%\def\ap#1#2#3{Ann. Phys. {\bf #1} (#2) #3}

%
\catcode`\@=11
\def\slash#1{\mathord{\mathpalette\c@ncel{#1}}}
\overfullrule=0pt

\def\underrel#1\over#2{\mathrel{\mathop{\kern\z@#1}\limits_{#2}}}

\catcode`\@=12

%%%%%%%%%%%%%%%%%%%%%%%%%%%%%%%%%%%%%%%%%%%%%%%%%%%%%%%%%%%%%%

%

%%%%%%%%%%%%%%%%%%%%%%%%%%%%%%%%%%%%%%%%%%%%%%%%%%%%%%%%%%%%%%
% new defs:

\def\[{[}
\def\]{]}

\def\comment#1{ }

%%%%%%%%%%%%%%%%%%%%%%%%%%%%%%%%%%%%%%%%%%%%%%%%%%%%%%%%%%%%%%
%%% Oskar's definitions:
%
%% A box for a short draft note
\def\draftnote#1{\ifdraft{\baselineskip2ex
                 \vbox{\kern1em\hrule\hbox{\vrule\kern1em\vbox{\kern1ex
                 \noindent \underbar{NOTE}: #1
             \vskip1ex}\kern1em\vrule}\hrule}}\fi}
%% A box for a short internal note
\def\internote#1{\ifinter{\baselineskip2ex
                 \vbox{\kern1em\hrule\hbox{\vrule\kern1em\vbox{\kern1ex
                 \noindent \underbar{Internal Note}: #1
             \vskip1ex}\kern1em\vrule}\hrule}}\fi}
%% A few internal words

%
%% Greek letters
%
%\def\al{\alpha}
%\def\bt{\beta}
%\def\gm{\gamma}                \def\Gm{\Gamma}
%\def\dl{\delta}                \def\Dl{\Delta}
%\def\ep{\epsilon}
%\def\vep{\varepsilon}

%\def\io{\iota}
%\def\kp{\kappa}
%\def\lm{\lambda}               \def\Lm{\Lambda}
%%\mu,\nu unchanged

%\def\th{\theta}               \def\Th{\Theta}
%\def\vth{\vartheta}
%%\phi unchanged               \Phi unchanged
%\def\vph{\varphi}
%%\psi unchanged               \Psi unchanged
%%\chi unchanged

%\def\om{\omega}               \def\Om{\Omega}
%%\pi unchanged                \Pi unchanged
%\def\vpi{\varpi}
%%\rho unchanged
%\def\vro{\varrho}
%\def\sg{\sigma}               \def\Sg{\Sigma}
%\def\vsg{\varsigma}
%%\tau unchanged
%\def\up{\upsilon}             \Up{\Upsilon}
%%\xi unchanged                \Xi unchanged
%%\eta unchanged
%\def\zt{\zeta}
%
%% Capital roman double letters (blackboard font)
%
%\def\inbar{\hskip.3em\vrule height1.5ex width.4pt depth0pt}
%\def\IC{\relax{\inbar\kern-.3em{\rm C}}}
%\def\IN{\relax{\rm I\kern-.16em N}}
%\def\IP{\relax{\rm I\kern-.18em P}}
%\def\IQ{\relax\hbox{$\inbar$\kern-.3em{\rm Q}}}
%\def\IR{\relax{\rm I\kern-.18em R}}
%\def\IZ{\relax{\rm Z\kern-.8em Z}}
%
%% Other Defs
%

%

\def\inv{^{-1}}

%mydef

\def\1{{\ds 1}}

% fraktur font
\newfam\frakfam
\font\teneufm=eufm10
\font\seveneufm=eufm7
\font\fiveeufm=eufm5
\textfont\frakfam=\teneufm
\scriptfont\frakfam=\seveneufm
\scriptscriptfont\frakfam=\fiveeufm
\def\frak{\fam\frakfam \teneufm}

\lref\NiarchosAH{
  V.~Niarchos,
  ``Seiberg dualities and the 3d/4d connection,''
JHEP {\bf 1207}, 075 (2012).
[arXiv:1205.2086 [hep-th]].
%%CITATION = arXiv:1205.2086%%
}

\lref\AharonyGP{
  O.~Aharony,
  ``IR duality in d = 3 N=2 supersymmetric USp( 2N(c)) and U(N(c)) gauge theories,''
Phys.\ Lett.\ B {\bf 404}, 71 (1997).
[hep-th/9703215].
%%CITATION = hep-th/9703215%%
}

%\AffleckAS
\lref\AffleckAS{
  I.~Affleck, J.~A.~Harvey and E.~Witten,
  ``Instantons and (Super)Symmetry Breaking in (2+1)-Dimensions,''
Nucl.\ Phys.\ B {\bf 206}, 413 (1982)..
%%CITATION = PRINT-82-0478 (PRINCETON)%%
}

%\IntriligatorID
\lref\IntriligatorID{
  K.~A.~Intriligator and N.~Seiberg,
  ``Duality, monopoles, dyons, confinement and oblique confinement in supersymmetric SO(  N(c)) gauge theories,''
Nucl.\ Phys.\ B {\bf 444}, 125 (1995).
[hep-th/9503179].
%%CITATION = hep-th/9503179%%
}

\lref\PasquettiFJ{
  S.~Pasquetti,
  ``Factorisation of N = 2 Theories on the Squashed 3-Sphere,''
JHEP {\bf 1204}, 120 (2012).
[arXiv:1111.6905 [hep-th]].
%%CITATION = arXiv:1111.6905%%
}

\lref\HarveyIT{
  J.~A.~Harvey,
 ``TASI 2003 lectures on anomalies,''
[hep-th/0509097].
%%CITATION = hep-th/0509097%%
}

\lref\BeemMB{
  C.~Beem, T.~Dimofte and S.~Pasquetti,
  ``Holomorphic Blocks in Three Dimensions,''
[arXiv:1211.1986 [hep-th]].
%%CITATION = arXiv:1211.1986%%
}

\lref\DiPietroBCA{
  L.~Di Pietro and Z.~Komargodski,
  ``Cardy formulae for SUSY theories in $d =$ 4 and $d =$ 6,''
JHEP {\bf 1412}, 031 (2014).
[arXiv:1407.6061 [hep-th]].
%%CITATION = arXiv:1407.6061%%
}

%\SeibergPQ
\lref\SeibergPQ{
  N.~Seiberg,
  ``Electric - magnetic duality in supersymmetric nonAbelian gauge theories,''
Nucl.\ Phys.\ B {\bf 435}, 129 (1995).
[hep-th/9411149].
%%CITATION = hep-th/9411149%%
}

%\AharonyBX
\lref\AharonyBX{
  O.~Aharony, A.~Hanany, K.~A.~Intriligator, N.~Seiberg and M.~J.~Strassler,
  ``Aspects of N=2 supersymmetric gauge theories in three-dimensions,''
Nucl.\ Phys.\ B {\bf 499}, 67 (1997).
[hep-th/9703110].
%%CITATION = hep-th/9703110%%
}

%\IntriligatorNE
\lref\IntriligatorNE{
  K.~A.~Intriligator and P.~Pouliot,
  ``Exact superpotentials, quantum vacua and duality in supersymmetric USp(N(c)) gauge theories,''
Phys.\ Lett.\ B {\bf 353}, 471 (1995).
[hep-th/9505006].
%%CITATION = hep-th/9505006%%
}

%\KarchUX
\lref\KarchUX{
  A.~Karch,
  ``Seiberg duality in three-dimensions,''
Phys.\ Lett.\ B {\bf 405}, 79 (1997).
[hep-th/9703172].
%%CITATION = hep-th/9703172%%
}

%\SafdiRE
\lref\SafdiRE{
  B.~R.~Safdi, I.~R.~Klebanov and J.~Lee,
  ``A Crack in the Conformal Window,''
[arXiv:1212.4502 [hep-th]].
%%CITATION = arXiv:1212.4502%%
}

\lref\SchweigertTG{
  C.~Schweigert,
  ``On moduli spaces of flat connections with nonsimply connected structure group,''
Nucl.\ Phys.\ B {\bf 492}, 743 (1997).
[hep-th/9611092].
%%CITATION = hep-th/9611092%%
}

%\GiveonZN
\lref\GiveonZN{
  A.~Giveon and D.~Kutasov,
  ``Seiberg Duality in Chern-Simons Theory,''
Nucl.\ Phys.\ B {\bf 812}, 1 (2009).
[arXiv:0808.0360 [hep-th]].
%%CITATION = arXiv:0808.0360%%
}

\lref\Spiridonov{
  Spiridonov, V.~P.,
  ``Aspects of elliptic hypergeometric functions,''
[arXiv:1307.2876 [math.CA]].
}

%\GaiottoBE
\lref\GaiottoBE{
  D.~Gaiotto, G.~W.~Moore and A.~Neitzke,
  ``Framed BPS States,''
[arXiv:1006.0146 [hep-th]].
%%CITATION = arXiv:1006.0146%%
}

\lref\AldayRS{
  L.~F.~Alday, M.~Bullimore and M.~Fluder,
  ``On S-duality of the Superconformal Index on Lens Spaces and 2d TQFT,''
JHEP {\bf 1305}, 122 (2013).
[arXiv:1301.7486 [hep-th]].
%%CITATION = arXiv:1301.7486%%
}

\lref\ArdehaliBLA{
  A.~Arabi Ardehali,
  ``High-temperature asymptotics of supersymmetric partition functions,''
JHEP {\bf 1607}, 025 (2016).
[arXiv:1512.03376 [hep-th]].
%%CITATION = MCTP-15-27%%
}

\lref\RazamatJXA{
  S.~S.~Razamat and M.~Yamazaki,
  ``S-duality and the N=2 Lens Space Index,''
[arXiv:1306.1543 [hep-th]].
%%CITATION = PUPT-2445%%
}

\lref\NiarchosAH{
  V.~Niarchos,
  ``Seiberg dualities and the 3d/4d connection,''
JHEP {\bf 1207}, 075 (2012).
[arXiv:1205.2086 [hep-th]].
%%CITATION = arXiv:1205.2086%%
}

\lref\almost{
  A.~Borel, R.~Friedman, J.~W.~Morgan,
  ``Almost commuting elements in compact Lie groups,''
arXiv:math/9907007.
%%CITATION = arXiv:1205.2086%%
}

\lref\BobevKZA{
  N.~Bobev, M.~Bullimore and H.~C.~Kim,
  ``Supersymmetric Casimir Energy and the Anomaly Polynomial,''
JHEP {\bf 1509}, 142 (2015).
[arXiv:1507.08553 [hep-th]].
%%CITATION = arXiv:1507.08553%%
}

\lref\KapustinJM{
  A.~Kapustin and B.~Willett,
  ``Generalized Superconformal Index for Three Dimensional Field Theories,''
[arXiv:1106.2484 [hep-th]].
%%CITATION = arXiv:1106.2484%%
}

\lref\AharonyGP{
  O.~Aharony,
  ``IR duality in d = 3 N=2 supersymmetric USp( 2N(c)) and U(N(c)) gauge theories,''
Phys.\ Lett.\ B {\bf 404}, 71 (1997).
[hep-th/9703215].
%%CITATION = hep-th/9703215%%
}

\lref\FestucciaWS{
  G.~Festuccia and N.~Seiberg,
  ``Rigid Supersymmetric Theories in Curved Superspace,''
JHEP {\bf 1106}, 114 (2011).
[arXiv:1105.0689 [hep-th]].
%%CITATION = arXiv:1105.0689%%
}

\lref\RomelsbergerEG{
  C.~Romelsberger,
  ``Counting chiral primaries in N = 1, d=4 superconformal field theories,''
Nucl.\ Phys.\ B {\bf 747}, 329 (2006).
[hep-th/0510060].
%%CITATION = hep-th/0510060%%
}

\lref\KapustinKZ{
  A.~Kapustin, B.~Willett and I.~Yaakov,
  ``Exact Results for Wilson Loops in Superconformal Chern-Simons Theories with Matter,''
JHEP {\bf 1003}, 089 (2010).
[arXiv:0909.4559 [hep-th]].
%%CITATION = arXiv:0909.4559%%
}

%\DolanQI
\lref\DolanQI{
  F.~A.~Dolan and H.~Osborn,
  ``Applications of the Superconformal Index for Protected Operators and q-Hypergeometric Identities to N=1 Dual Theories,''
Nucl.\ Phys.\ B {\bf 818}, 137 (2009).
[arXiv:0801.4947 [hep-th]].
%%CITATION = arXiv:0801.4947%%
}

\lref\GaddeIA{
  A.~Gadde and W.~Yan,
  ``Reducing the 4d Index to the $S^3$ Partition Function,''
JHEP {\bf 1212}, 003 (2012).
[arXiv:1104.2592 [hep-th]].
%%CITATION = arXiv:1104.2592%%
}

\lref\DolanRP{
  F.~A.~H.~Dolan, V.~P.~Spiridonov and G.~S.~Vartanov,
  ``From 4d superconformal indices to 3d partition functions,''
Phys.\ Lett.\ B {\bf 704}, 234 (2011).
[arXiv:1104.1787 [hep-th]].
%%CITATION = arXiv:1104.1787%%
}

\lref\ImamuraUW{
  Y.~Imamura,
 ``Relation between the 4d superconformal index and the $S^3$ partition function,''
JHEP {\bf 1109}, 133 (2011).
[arXiv:1104.4482 [hep-th]].
%%CITATION = arXiv:1104.4482%%
}

\lref\BeemYN{
  C.~Beem and A.~Gadde,
  ``The $N=1$ superconformal index for class $S$ fixed points,''
JHEP {\bf 1404}, 036 (2014).
[arXiv:1212.1467 [hep-th]].
%%CITATION = arXiv:1212.1467%%
}

\lref\HamaEA{
  N.~Hama, K.~Hosomichi and S.~Lee,
  ``SUSY Gauge Theories on Squashed Three-Spheres,''
JHEP {\bf 1105}, 014 (2011).
[arXiv:1102.4716 [hep-th]].
%%CITATION = arXiv:1102.4716%%
}

\lref\GaddeEN{
  A.~Gadde, L.~Rastelli, S.~S.~Razamat and W.~Yan,
  ``On the Superconformal Index of N=1 IR Fixed Points: A Holographic Check,''
JHEP {\bf 1103}, 041 (2011).
[arXiv:1011.5278 [hep-th]].
%%CITATION = arXiv:1011.5278%%
}

\lref\EagerHX{
  R.~Eager, J.~Schmude and Y.~Tachikawa,
  ``Superconformal Indices, Sasaki-Einstein Manifolds, and Cyclic Homologies,''
[arXiv:1207.0573 [hep-th]].
%%CITATION = arXiv:1207.0573%%
}

%\AffleckAS
\lref\AffleckAS{
  I.~Affleck, J.~A.~Harvey and E.~Witten,
  ``Instantons and (Super)Symmetry Breaking in (2+1)-Dimensions,''
Nucl.\ Phys.\ B {\bf 206}, 413 (1982)..
%%CITATION = PRINT-82-0478 (PRINCETON)%%
}

\lref\BahGPH{
  I.~Bah, A.~Hanany, K.~Maruyoshi, S.~S.~Razamat, Y.~Tachikawa and G.~Zafrir,
  ``4d $ {\cal N}=1 $ from 6d $ {\cal N}=\left(1,0\right) $ on a torus with fluxes,''
JHEP {\bf 1706}, 022 (2017).
[arXiv:1702.04740 [hep-th]].
%%CITATION = IPMU17-0013%%
}

%\SeibergPQ
\lref\SeibergPQ{
  N.~Seiberg,
  ``Electric - magnetic duality in supersymmetric nonAbelian gauge theories,''
Nucl.\ Phys.\ B {\bf 435}, 129 (1995).
[hep-th/9411149].
%%CITATION = hep-th/9411149%%
}

\lref\BahDG{
  I.~Bah, C.~Beem, N.~Bobev and B.~Wecht,
  ``Four-Dimensional SCFTs from M5-Branes,''
JHEP {\bf 1206}, 005 (2012).
[arXiv:1203.0303 [hep-th]].
%%CITATION = arXiv:1203.0303%%
}

\lref\debult{
  F.~van~de~Bult,
  ``Hyperbolic Hypergeometric Functions,''
University of Amsterdam Ph.D. thesis
}

\lref\OhmoriAMP{
  K.~Ohmori, H.~Shimizu, Y.~Tachikawa and K.~Yonekura,
  ``Anomaly polynomial of general $6d$ SCFTs,''
PTEP {\bf 2014}, 103B07 (2014).
[arXiv:1408.5572 [hep-th]].}

\lref\Shamirthesis{
  I.~Shamir,
  ``Aspects of three dimensional Seiberg duality,''
  M. Sc. thesis submitted to the Weizmann Institute of Science, April 2010.
  }

\lref\slthreeZ{
  J.~Felder, A.~Varchenko,
  ``The elliptic gamma function and $SL(3,Z) \times Z^3$,'' $\;\;$
[arXiv:math/0001184].
}

\lref\BeniniNC{
  F.~Benini, T.~Nishioka and M.~Yamazaki,
  ``4d Index to 3d Index and 2d TQFT,''
Phys.\ Rev.\ D {\bf 86}, 065015 (2012).
[arXiv:1109.0283 [hep-th]].
%%CITATION = arXiv:1109.0283%%
}

\lref\GaiottoWE{
  D.~Gaiotto,
  ``N=2 dualities,''
  JHEP {\bf 1208}, 034 (2012).
  [arXiv:0904.2715 [hep-th]].
  %%CITATION = arXiv:0904.2715%%
}

\lref\SpiridonovZA{
  V.~P.~Spiridonov and G.~S.~Vartanov,
  ``Elliptic Hypergeometry of Supersymmetric Dualities,''
Commun.\ Math.\ Phys.\  {\bf 304}, 797 (2011).
[arXiv:0910.5944 [hep-th]].
%%CITATION = arXiv:0910.5944%%
}

%\BeniniMF
\lref\BeniniMF{
  F.~Benini, C.~Closset and S.~Cremonesi,
  ``Comments on 3d Seiberg-like dualities,''
JHEP {\bf 1110}, 075 (2011).
[arXiv:1108.5373 [hep-th]].
%%CITATION = arXiv:1108.5373%%
}

\lref\BeniniGI{
  F.~Benini, S.~Benvenuti and Y.~Tachikawa,
  ``Webs of five-branes and N=2 superconformal field theories,''
JHEP {\bf 0909}, 052 (2009).
[arXiv:0906.0359 [hep-th]].
%%CITATION = arXiv:0906.0359%%
}

%\ClossetVP
\lref\ClossetVP{
  C.~Closset, T.~T.~Dumitrescu, G.~Festuccia, Z.~Komargodski and N.~Seiberg,
  ``Comments on Chern-Simons Contact Terms in Three Dimensions,''
JHEP {\bf 1209}, 091 (2012).
[arXiv:1206.5218 [hep-th]].
%%CITATION = arXiv:1206.5218%%
}

\lref\SpiridonovHF{
  V.~P.~Spiridonov and G.~S.~Vartanov,
  ``Elliptic hypergeometry of supersymmetric dualities II. Orthogonal groups, knots, and vortices,''
[arXiv:1107.5788 [hep-th]].
%%CITATION = arXiv:1107.5788%%
}

\lref\RazamatQFA{
  S.~S.~Razamat,
  ``On the $\cal{N} =$ 2 superconformal index and eigenfunctions of the elliptic RS model,''
Lett.\ Math.\ Phys.\  {\bf 104}, 673 (2014).
[arXiv:1309.0278 [hep-th]].
}

\lref\SpiridonovWW{
  V.~P.~Spiridonov and G.~S.~Vartanov,
  ``Elliptic hypergeometric integrals and 't Hooft anomaly matching conditions,''
JHEP {\bf 1206}, 016 (2012).
[arXiv:1203.5677 [hep-th]].
%%CITATION = arXiv:1203.5677%%
}

\lref\HeckmanXDL{
  J.~J.~Heckman, P.~Jefferson, T.~Rudelius and C.~Vafa,
  ``Punctures for Theories of Class ${\cal{S}}_\Gamma$,''
[arXiv:1609.01281 [hep-th]].
%%CITATION = arXiv:1609.01281%%
}

\lref\DimoftePY{
  T.~Dimofte, D.~Gaiotto and S.~Gukov,
  ``3-Manifolds and 3d Indices,''
[arXiv:1112.5179 [hep-th]].
%%CITATION = arXiv:1112.5179%%
}

\lref\KimWB{
  S.~Kim,
  ``The Complete superconformal index for N=6 Chern-Simons theory,''
Nucl.\ Phys.\ B {\bf 821}, 241 (2009), [Erratum-ibid.\ B {\bf 864}, 884 (2012)].
[arXiv:0903.4172 [hep-th]].
%%CITATION = arXiv:0903.4172%%
}

%\GP
\lref\WillettGP{
  B.~Willett and I.~Yaakov,
  ``N=2 Dualities and Z Extremization in Three Dimensions,''
[arXiv:1104.0487 [hep-th]].
%%CITATION = arXiv:1104.0487%%
}

\lref\ImamuraSU{
  Y.~Imamura and S.~Yokoyama,
  ``Index for three dimensional superconformal field theories with general R-charge assignments,''
JHEP {\bf 1104}, 007 (2011).
[arXiv:1101.0557 [hep-th]].
%%CITATION = arXiv:1101.0557%%
}

%\FreedYA
\lref\FreedYA{
  D.~S.~Freed, G.~W.~Moore and G.~Segal,
  ``The Uncertainty of Fluxes,''
Commun.\ Math.\ Phys.\  {\bf 271}, 247 (2007).
[hep-th/0605198].
%%CITATION = hep-th/0605198%%
}

\lref\HwangQT{
  C.~Hwang, H.~Kim, K.~-J.~Park and J.~Park,
  ``Index computation for 3d Chern-Simons matter theory: test of Seiberg-like duality,''
JHEP {\bf 1109}, 037 (2011).
[arXiv:1107.4942 [hep-th]].
%%CITATION = arXiv:1107.4942%%
}

\lref\GreenDA{
  D.~Green, Z.~Komargodski, N.~Seiberg, Y.~Tachikawa and B.~Wecht,
  ``Exactly Marginal Deformations and Global Symmetries,''
JHEP {\bf 1006}, 106 (2010).
[arXiv:1005.3546 [hep-th]].
%%CITATION = arXiv:1005.3546%%
}

\lref\IntriligatorJJ{
  K.~A.~Intriligator and B.~Wecht,
  ``The Exact superconformal R symmetry maximizes a,''
Nucl.\ Phys.\ B {\bf 667}, 183 (2003).
[hep-th/0304128].
%%CITATION = hep-th/0304128%%
}

\lref\RazamatDPL{
  S.~S.~Razamat, C.~Vafa and G.~Zafrir,
  ``4d $ {  
  \cal N}=1 $ from 6d (1, 0),''
JHEP {\bf 1704}, 064 (2017).
[arXiv:1610.09178 [hep-th]].
%%CITATION = arXiv:1610.09178%%
}

%\IntriligatorID
\lref\IntriligatorID{
  K.~A.~Intriligator and N.~Seiberg,
  ``Duality, monopoles, dyons, confinement and oblique confinement in supersymmetric SO(  N(c)) gauge theories,''
Nucl.\ Phys.\ B {\bf 444}, 125 (1995).
[hep-th/9503179].
%%CITATION = hep-th/9503179%%
}

%\SeibergNZ
\lref\SeibergNZ{
  N.~Seiberg and E.~Witten,
  ``Gauge dynamics and compactification to three-dimensions,''
In *Saclay 1996, The mathematical beauty of physics* 333-366.
[hep-th/9607163].
%%CITATION = hep-th/9607163%%
}

\lref\KinneyEJ{
  J.~Kinney, J.~M.~Maldacena, S.~Minwalla and S.~Raju,
  ``An Index for 4 dimensional super conformal theories,''
  Commun.\ Math.\ Phys.\  {\bf 275}, 209 (2007).
  [hep-th/0510251].
 %%CITATION = hep-th/0510251%%
}

\lref\NakayamaUR{
  Y.~Nakayama,
  ``Index for supergravity on AdS(5) x T**1,1 and conifold gauge theory,''
Nucl.\ Phys.\ B {\bf 755}, 295 (2006).
[hep-th/0602284].
%%CITATION = hep-th/0602284%%
}

\lref\GaddeKB{
  A.~Gadde, E.~Pomoni, L.~Rastelli and S.~S.~Razamat,
  ``S-duality and 2d Topological QFT,''
JHEP {\bf 1003}, 032 (2010).
[arXiv:0910.2225 [hep-th]].
}

\lref\GaddeTE{
  A.~Gadde, L.~Rastelli, S.~S.~Razamat and W.~Yan,
  ``The Superconformal Index of the $E_6$ SCFT,''
JHEP {\bf 1008}, 107 (2010).
[arXiv:1003.4244 [hep-th]].
%%CITATION = arXiv:1003.4244%%
}

%\AharonyCI
\lref\AharonyCI{
  O.~Aharony and I.~Shamir,
  ``On $O(N_c)$ d=3 N=2 supersymmetric QCD Theories,''
JHEP {\bf 1112}, 043 (2011).
[arXiv:1109.5081 [hep-th]].
%%CITATION = arXiv:1109.5081%%
}

%\GiveonSR
\lref\GiveonSR{
  A.~Giveon and D.~Kutasov,
  ``Brane dynamics and gauge theory,''
Rev.\ Mod.\ Phys.\  {\bf 71}, 983 (1999).
[hep-th/9802067].
%%CITATION = hep-th/9802067%%
}

\lref\SpiridonovQV{
  V.~P.~Spiridonov and G.~S.~Vartanov,
  ``Superconformal indices of ${\cal N}=4$ SYM field theories,''
Lett.\ Math.\ Phys.\  {\bf 100}, 97 (2012).
[arXiv:1005.4196 [hep-th]].
%%CITATION = arXiv:1005.4196%%
}
\lref\GaddeUV{
  A.~Gadde, L.~Rastelli, S.~S.~Razamat and W.~Yan,
  ``Gauge Theories and Macdonald Polynomials,''
Commun.\ Math.\ Phys.\  {\bf 319}, 147 (2013).
[arXiv:1110.3740 [hep-th]].
}

\lref\AlvarezGaumeIG{
  L.~Alvarez-Gaume and E.~Witten,
 ``Gravitational Anomalies,''
Nucl.\ Phys.\ B {\bf 234}, 269 (1984).
%%CITATION = HUTP-83/A039%%
}

%\KapustinGH
\lref\KapustinGH{
  A.~Kapustin,
  ``Seiberg-like duality in three dimensions for orthogonal gauge groups,''
[arXiv:1104.0466 [hep-th]].
%%CITATION = arXiv:1104.0466%%
}

\lref\orthogpaper{O. Aharony, S. S. Razamat, N.~Seiberg and B.~Willett, 
``3d dualities from 4d dualities for orthogonal groups,''
[arXiv:1307.0511 [hep-th]].
%%CITATION = WIS-03-13-APR-DPPA%%
}

\lref\readinglines{
  O.~Aharony, N.~Seiberg and Y.~Tachikawa,
  ``Reading between the lines of four-dimensional gauge theories,''
[arXiv:1305.0318 [hep-th]].
%%CITATION = WIS-03-13-APR-DPPA%%
}

%\WittenNV
\lref\WittenNV{
  E.~Witten,
  ``Supersymmetric index in four-dimensional gauge theories,''
Adv.\ Theor.\ Math.\ Phys.\  {\bf 5}, 841 (2002).
[hep-th/0006010].
%%CITATION = hep-th/0006010%%
}

\lref\GaddeUV{
  A.~Gadde, L.~Rastelli, S.~S.~Razamat and W.~Yan,
  ``Gauge Theories and Macdonald Polynomials,''
Commun.\ Math.\ Phys.\  {\bf 319}, 147 (2013).
[arXiv:1110.3740 [hep-th]].
%%CITATION = arXiv:1110.3740%%
}

\lref\GaddeIK{
  A.~Gadde, L.~Rastelli, S.~S.~Razamat and W.~Yan,
  ``The 4d Superconformal Index from q-deformed 2d Yang-Mills,''
Phys.\ Rev.\ Lett.\  {\bf 106}, 241602 (2011).
[arXiv:1104.3850 [hep-th]].
%%CITATION = arXiv:1104.3850%%
}

\lref\GaiottoXA{
  D.~Gaiotto, L.~Rastelli and S.~S.~Razamat,
  ``Bootstrapping the superconformal index with surface defects,''
JHEP {\bf 1301}, 022 (2013).
[arXiv:1207.3577 [hep-th]].
%%CITATION = arXiv:1207.3577%%
}

\lref\GaiottoUQ{
  D.~Gaiotto and S.~S.~Razamat,
  ``Exceptional Indices,''
JHEP {\bf 1205}, 145 (2012).
[arXiv:1203.5517 [hep-th]].
%%CITATION = arXiv:1203.5517%%
}

\lref\RazamatUV{
  S.~S.~Razamat,
  ``On a modular property of N=2 superconformal theories in four dimensions,''
JHEP {\bf 1210}, 191 (2012).
[arXiv:1208.5056 [hep-th]].
%%CITATION = arXiv:1208.5056%%
}

\lref\noumi{
  Y.~Komori, M.~Noumi, J.~Shiraishi,
  ``Kernel Functions for Difference Operators of Ruijsenaars Type and Their Applications,''
SIGMA 5 (2009), 054.
[arXiv:0812.0279 [math.QA]].
%%CITATION = arXiv:1208.5056%%
}

\lref\SpirWarnaar{
  V.~P.~Spiridonov and S.~O.~Warnaar,
  ``Inversions of integral operators and elliptic beta integrals on root systems,''
Adv. Math. 207 (2006), 91-132
[arXiv:math/0411044].
}

\lref\RazamatJXA{
  S.~S.~Razamat and M.~Yamazaki,
  ``S-duality and the N=2 Lens Space Index,''
[arXiv:1306.1543 [hep-th]].
%%CITATION = PUPT-2445%%
}

\lref\RazamatOPA{
  S.~S.~Razamat and B.~Willett,
  ``Global Properties of Supersymmetric Theories and the Lens Space,''
[arXiv:1307.4381 [hep-th]].
%%CITATION = arXiv:1307.4381%%
}

\lref\GaddeTE{
  A.~Gadde, L.~Rastelli, S.~S.~Razamat and W.~Yan,
  ``The Superconformal Index of the $E_6$ SCFT,''
JHEP {\bf 1008}, 107 (2010).
[arXiv:1003.4244 [hep-th]].
%%CITATION = arXiv:1003.4244%%
}

\lref\deBult{
  F.~J.~van~de~Bult,
  ``An elliptic hypergeometric integral with $W(F_4)$ symmetry,''
The Ramanujan Journal, Volume 25, Issue 1 (2011)
[arXiv:0909.4793[math.CA]].
%%CITATION = arXiv:1003.4244%%
}

\lref\MaruyoshiCAF{
  K.~Maruyoshi and J.~Yagi,
  ``Surface defects as transfer matrices,''
[arXiv:1606.01041 [hep-th]].
%%CITATION = IMPERIAL-TP-16-KM-01%%
}

\lref\GaddeKB{
  A.~Gadde, E.~Pomoni, L.~Rastelli and S.~S.~Razamat,
  ``S-duality and 2d Topological QFT,''
JHEP {\bf 1003}, 032 (2010).
[arXiv:0910.2225 [hep-th]].
%%CITATION = arXiv:0910.2225%%
}

\lref\ArgyresCN{
  P.~C.~Argyres and N.~Seiberg,
  ``S-duality in N=2 supersymmetric gauge theories,''
JHEP {\bf 0712}, 088 (2007).
[arXiv:0711.0054 [hep-th]].
%%CITATION = arXiv:0711.0054%%
}

\lref\SpirWarnaar{
  V.~P.~Spiridonov and S.~O.~Warnaar,
  ``Inversions of integral operators and elliptic beta integrals on root systems,''
Adv. Math. 207 (2006), 91-132
[arXiv:math/0411044].
%%CITATION = arXiv:1003.4244%%
}

\lref\GaiottoHG{
  D.~Gaiotto, G.~W.~Moore and A.~Neitzke,
  ``Wall-crossing, Hitchin Systems, and the WKB Approximation,''
[arXiv:0907.3987 [hep-th]].
%%CITATION = arXiv:0907.3987%%
}

\lref\RuijsenaarsVQ{
  S.~N.~M.~Ruijsenaars and H.~Schneider,
  ``A New Class Of Integrable Systems And Its Relation To Solitons,''
Annals Phys.\  {\bf 170}, 370 (1986).
}

\lref\RuijsenaarsPP{
  S.~N.~M.~Ruijsenaars,
  ``Complete Integrability Of Relativistic Calogero-moser Systems And Elliptic Function Identities,''
Commun.\ Math.\ Phys.\  {\bf 110}, 191 (1987).
%%CITATION = MPI-PAE/PTh 26/86%%
}

\lref\HallnasNB{
  M.~Hallnas and S.~Ruijsenaars,
  ``Kernel functions and Baecklund transformations for relativistic Calogero-Moser and Toda systems,''
J.\ Math.\ Phys.\  {\bf 53}, 123512 (2012).
}

\lref\kernelA{
S.~Ruijsenaars,
  ``Elliptic integrable systems of Calogero-Moser type: Some new results on joint eigenfunctions'', in Proceedings of the 2004 Kyoto Workshop on "Elliptic integrable systems", (M. Noumi, K. Takasaki, Eds.), Rokko Lectures in Math., no. 18, Dept. of Math., Kobe Univ.
}

\lref\ellRSreview{
Y.~Komori and S.~Ruijsenaars,
  ``Elliptic integrable systems of Calogero-Moser type: A survey'', in Proceedings of the 2004 Kyoto Workshop on "Elliptic integrable systems", (M. Noumi, K. Takasaki, Eds.), Rokko Lectures in Math., no. 18, Dept. of Math., Kobe Univ.
}

\lref\langmann{
E.~Langmann,
  ``An explicit solution of the (quantum) elliptic Calogero-Sutherland model'', [arXiv:math-ph/0407050].
}

\lref\TachikawaWI{
  Y.~Tachikawa,
  ``4d partition function on $S^1 \times S^3$ and 2d Yang-Mills with nonzero area,''
PTEP {\bf 2013}, 013B01 (2013).
[arXiv:1207.3497 [hep-th]].
%%CITATION = arXiv:1207.3497%%
}

\lref\MinahanFG{
  J.~A.~Minahan and D.~Nemeschansky,
  ``An N=2 superconformal fixed point with E(6) global symmetry,''
Nucl.\ Phys.\ B {\bf 482}, 142 (1996).
[hep-th/9608047].
%%CITATION = hep-th/9608047%%
}

\lref\AldayKDA{
  L.~F.~Alday, M.~Bullimore, M.~Fluder and L.~Hollands,
  ``Surface defects, the superconformal index and q-deformed Yang-Mills,''
[arXiv:1303.4460 [hep-th]].
%%CITATION = arXiv:1303.4460%%
}

\lref\FukudaJR{
  Y.~Fukuda, T.~Kawano and N.~Matsumiya,
  ``5D SYM and 2D q-Deformed YM,''
Nucl.\ Phys.\ B {\bf 869}, 493 (2013).
[arXiv:1210.2855 [hep-th]].
%%CITATION = arXiv:1210.2855%%
}

\lref\XieHS{
  D.~Xie,
  ``General Argyres-Douglas Theory,''
JHEP {\bf 1301}, 100 (2013).
[arXiv:1204.2270 [hep-th]].
%%CITATION = arXiv:1204.2270%%
}

\lref\DrukkerSR{
  N.~Drukker, T.~Okuda and F.~Passerini,
  ``Exact results for vortex loop operators in 3d supersymmetric theories,''
[arXiv:1211.3409 [hep-th]].
%%CITATION = arXiv:1211.3409%%
}
\lref\XieGMA{
  D.~Xie,
  ``M5 brane and four dimensional N = 1 theories I,''
JHEP {\bf 1404}, 154 (2014).
[arXiv:1307.5877 [hep-th]].
%%CITATION = arXiv:1307.5877%%
}

\lref\qinteg{
  M.~Rahman, A.~Verma,
  ``A q-integral representation of Rogers' q-ultraspherical polynomials and some applications,''
Constructive Approximation
1986, Volume 2, Issue 1.
}

\lref\qintegOK{
  A.~Okounkov,
  ``(Shifted) Macdonald Polynomials: q-Integral Representation and Combinatorial Formula,''
Compositio Mathematica
June 1998, Volume 112, Issue 2. 
[arXiv:q-alg/9605013].
}

\lref\macNest{
 H.~Awata, S.~Odake, J.~Shiraishi,
  ``Integral Representations of the Macdonald Symmetric Functions,''
Commun. Math. Phys. 179 (1996) 647.
[arXiv:q-alg/9506006].
}

\lref\deBult{
  F.~J.~van~de~Bult,
  ``An elliptic hypergeometric integral with $W(F_4)$ symmetry,''
The Ramanujan Journal, Volume 25, Issue 1 (2011)
[arXiv:0909.4793[math.CA]].
}

\lref\Rains{
E.~M.~Rains,
  ``Transformations of elliptic hypergometric integrals,''
Annals of Mathematics, Volume  171, Issue 1 (2010)
[arXiv:math/0309252].
}

\lref\ItoFPL{
  Y.~Ito and Y.~Yoshida,
  ``Superconformal index with surface defects for class ${\cal S}_k$,''
[arXiv:1606.01653 [hep-th]].
%%CITATION = KIAS-P16040%%
}

\lref\BeniniMZ{
  F.~Benini, Y.~Tachikawa and B.~Wecht,
  ``Sicilian gauge theories and N=1 dualities,''
JHEP {\bf 1001}, 088 (2010).
[arXiv:0909.1327 [hep-th]].
%%CITATION = arXiv:0909.1327%%
}

\lref\DimoftePD{
  T.~Dimofte and D.~Gaiotto,
  ``An E7 Surprise,''
JHEP {\bf 1210}, 129 (2012).
[arXiv:1209.1404 [hep-th]].
}

\lref\GaddeFMA{
  A.~Gadde, K.~Maruyoshi, Y.~Tachikawa and W.~Yan,
  ``New N=1 Dualities,''
JHEP {\bf 1306}, 056 (2013).
[arXiv:1303.0836 [hep-th]].
}

\lref\AgarwalVLA{
  P.~Agarwal, K.~Intriligator and J.~Song,
  ``Infinitely many ${\cal N}=1 $ dualities from m + 1 - m = 1,''
JHEP {\bf 1510}, 035 (2015).
[arXiv:1505.00255 [hep-th]].
}

\lref\GorskyTN{
  A.~Gorsky,
  ``Dualities in integrable systems and N=2 SUSY theories,''
J.\ Phys.\ A {\bf 34}, 2389 (2001).
[hep-th/9911037].
%%CITATION = hep-th/9911037%%
}

\lref\FockAE{
  V.~Fock, A.~Gorsky, N.~Nekrasov and V.~Rubtsov,
  ``Duality in integrable systems and gauge theories,''
JHEP {\bf 0007}, 028 (2000).
[hep-th/9906235].
%%CITATION = hep-th/9906235%%
}

\lref\RazamatPTA{
  S.~S.~Razamat and B.~Willett,
  ``Down the rabbit hole with theories of class $ \cal S $,''
JHEP {\bf 1410}, 99 (2014).
[arXiv:1403.6107 [hep-th]].
}

\lref\RazamatL{ A.~Gadde, S.~S.~Razamat, and  B.~Willett,
  ``A ``Lagrangian'' for a non-Lagrangian theory,''
  Phys. Rev. Lett. {\bf 115}, 171604 (2015).
  [arXiv:1505.05834 [hep-th]].
  }
  
  \lref\OhmoriPUA{
  K.~Ohmori, H.~Shimizu, Y.~Tachikawa and K.~Yonekura,
  ``6d ${\cal N}=(1,0)$ theories on $T^2$ and class S theories: Part I,''
JHEP {\bf 1507}, 014 (2015).
[arXiv:1503.06217 [hep-th]].
}

\lref\OhmoriPIA{
  K.~Ohmori, H.~Shimizu, Y.~Tachikawa and K.~Yonekura,
  ``6d ${\cal N}=(1,\;0) $ theories on S$^{1}$ /T$^{2}$ and class S theories: part II,''
JHEP {\bf 1512}, 131 (2015).
[arXiv:1508.00915 [hep-th]].
}

\lref\DelZottoRCA{
  M.~Del Zotto, C.~Vafa and D.~Xie,
  ``Geometric engineering, mirror symmetry and $ 6{{d}}_{\left(1,0\right)}\to 4{{d}}_{\left({\cal N}=2\right)} $,''
JHEP {\bf 1511}, 123 (2015).
[arXiv:1504.08348 [hep-th]].
}

\lref\HananyPFA{
  A.~Hanany and K.~Maruyoshi,
  ``Chiral theories of class $ {\cal S} $,''
JHEP {\bf 1512}, 080 (2015).
[arXiv:1505.05053 [hep-th]].
}

\lref\CsakiEU{
  C.~Csaki, W.~Skiba and M.~Schmaltz,
  ``Exact results and duality for SP(2N) SUSY gauge theories with an antisymmetric tensor,''
Nucl.\ Phys.\ B {\bf 487}, 128 (1997).
[hep-th/9607210].
%%CITATION = hep-th/9607210%%
}

\lref\GaiottoLCA{
  D.~Gaiotto and A.~Tomasiello,
  ``Holography for (1,0) theories in six dimensions,''
JHEP {\bf 1412}, 003 (2014).
[arXiv:1404.0711 [hep-th]].
%%CITATION = arXiv:1404.0711%%
}

\lref\BenvenutiKUD{
  S.~Benvenuti and S.~Giacomelli,
  ``Abelianization and Sequential Confinement in $2+1$ dimensions,''
[arXiv:1706.04949 [hep-th]].
%%CITATION = arXiv:1706.04949%%
}

\lref\skm{
H. C.  Kim, S. S. Razamat, C. Vafa, G. Zafrir,    ``E-string theory on Riemann surfaces,'' [arXiv:1709:02496].
}

\lref\ApruzziZNA{
  F.~Apruzzi, M.~Fazzi, A.~Passias and A.~Tomasiello,
  ``Supersymmetric AdS$_{5}$ solutions of massive IIA supergravity,''
JHEP {\bf 1506}, 195 (2015).
[arXiv:1502.06620 [hep-th]].
%%CITATION = arXiv:1502.06620%%
}

\lref\BenvenutiLLE{
  S.~Benvenuti and S.~Giacomelli,
  ``Compactification of dualities with decoupled operators and $3d$ mirror symmetry,''
[arXiv:1706.02225 [hep-th]].
%%CITATION = arXiv:1706.02225%%
}

\lref\IntriligatorAX{
  K.~A.~Intriligator, R.~G.~Leigh and M.~J.~Strassler,
 ``New examples of duality in chiral and nonchiral supersymmetric gauge theories,''
Nucl.\ Phys.\ B {\bf 456}, 567 (1995).
[hep-th/9506148].
%%CITATION = hep-th/9506148%%
}

\lref\ComanBQQ{
  I.~Coman, E.~Pomoni, M.~Taki and F.~Yagi,
  ``Spectral curves of ${\cal N}=1$ theories of class ${\cal S}_k$,''
[arXiv:1512.06079 [hep-th]].
}

\lref\IntriligatorFF{
  K.~A.~Intriligator,
  ``New RG fixed points and duality in supersymmetric SP(N(c)) and SO(N(c)) gauge theories,''
Nucl.\ Phys.\ B {\bf 448}, 187 (1995).
[hep-th/9505051].
%%CITATION = hep-th/9505051%%
}

\lref\FrancoJNA{
  S.~Franco, H.~Hayashi and A.~Uranga,
  ``Charting Class ${\cal S}_k$ Territory,''
Phys.\ Rev.\ D {\bf 92}, no. 4, 045004 (2015).
[arXiv:1504.05988 [hep-th]].
}

\lref\SpiridonovZR{
  V.~P.~Spiridonov and G.~S.~Vartanov,
 ``Superconformal indices for N = 1 theories with multiple duals,''
Nucl.\ Phys.\ B {\bf 824}, 192 (2010).
[arXiv:0811.1909 [hep-th]].
}

\lref\AharonyDHA{
  O.~Aharony, S.~S.~Razamat, N.~Seiberg and B.~Willett,
  ``3d dualities from 4d dualities,''
JHEP {\bf 1307}, 149 (2013).
[arXiv:1305.3924 [hep-th]].}

\lref\HeckmanPVA{
  J.~J.~Heckman, D.~R.~Morrison and C.~Vafa,
  ``On the Classification of 6D SCFTs and Generalized ADE Orbifolds,''
JHEP {\bf 1405}, 028 (2014), Erratum: [JHEP {\bf 1506}, 017 (2015)].
[arXiv:1312.5746 [hep-th]].
%%CITATION = arXiv:1312.5746%%
}

\lref\CsakiCU{
  C.~Csaki, M.~Schmaltz, W.~Skiba and J.~Terning,
  ``Selfdual N=1 SUSY gauge theories,''
Phys.\ Rev.\ D {\bf 56}, 1228 (1997).
[hep-th/9701191].
%%CITATION = hep-th/9701191%%
}

\lref\DelZottoHPA{
  M.~Del Zotto, J.~J.~Heckman, A.~Tomasiello and C.~Vafa,
  ``6d Conformal Matter,''
JHEP {\bf 1502}, 054 (2015).
[arXiv:1407.6359 [hep-th]].
%%CITATION = arXiv:1407.6359%%
}

\lref\GaiottoUSA{
  D.~Gaiotto and S.~S.~Razamat,
  ``${\cal N}=1 $ theories of class $ {\cal S}_k $,''
JHEP {\bf 1507}, 073 (2015).
[arXiv:1503.05159 [hep-th]].
}

\Title{\vbox{\baselineskip12pt
}}
{\vbox{
\centerline{$E_8$ orbits of IR dualities}
\vskip7pt 
\centerline{}
}
}

\centerline{Shlomo S. Razamat$^a$ and Gabi Zafrir$^b$}
\bigskip
\centerline{{\it $^a$Physics Department, Technion, Haifa, Israel 32000}}

\centerline{  
{\it $^b$Kavli IPMU (WPI), UTIAS, the University of Tokyo, Kashiwa, Chiba 277-8583, Japan
}}
\bigskip
\centerline{{\it }}
\vskip.1in \vskip.2in \centerline{\bf Abstract}

We discuss $USp( 2n)$ supersymmetric models with eight fundamental fields and a field in the antisymmetric representation. Turning on the most generic superpotentials, coupling pairs of fundamental fields to powers of the antisymmetric field while preserving an R symmetry, we give evidence for the statement that the models are connected by a large network of dualities which can be organized into orbits of the Weyl group of $E_8$. We make also several curious observations about such models. In particular, we argue that a $USp(2m)$ model with the addition of singlet fields and even rank $m$  flows in the IR to a CFT with $E_7\times U(1)$ symmetry. We also discuss an infinite number of duals for the $USp(2)$ theory with eight fundamentals and no superpotential.

\vskip.2in

\noindent  

\vfill

\Date{October 2017}

\newsec{Introduction}

Different QFTs can flow to the same conformal field theory in the IR. This phenomenon is  usually referred to as IR duality in high energy physics or universality in statistical physics.  Considering supersymmetric theories in four dimensions following \SeibergPQ, 
one can discover various examples of seemingly very different models, for example possessing different gauge groups, which nevertheless reside in the same universality class. An understanding why two different looking theories in the UV flow to the same fixed point is rather lacking at 
the moment.\foot{There is though growing evidence that such dualities can be understood  by constructing dual four dimensional theories  as geometrically equivalent but different looking  compactifications of a six dimensional model (see for examples \refs{\GaiottoWE,\BahDG, \AgarwalVLA, \GaiottoUSA, \RazamatDPL,\BahGPH,\skm}).} An interesting question one can ask to aid such an understanding is whether there is any  structure relating different theories in a certain universality class.

In this short note we will discuss such a structure in a very particular setup. We consider $USp(2m)$ gauge theories with eight fundamental chiral fields $Q_i$,  a field in the antisymmetric representation $X$, and possibly a superpotential with gauge singlet fields. It so happens that these models are interrelated by a large network of dualities and this network has intriguing group theoretic structure. In particular there are dualities relating  models with fixed rank but different superpotentials and a non trivial map of operators/symmetries between various sides of the duality. This duality web forms
\refs{\Rains,\SpiridonovZR}
  orbits of the Weyl group of $E_7$. Here we discuss yet another duality transformation which relates theories with different rank. In particular a $USp(2m)$ model with superpotential $Q_2 Q_1 X^n$ is in the same universality class as a $USp(2n)$ model with superpotential $Q_2 Q_1 X^m$ when certain singlet fields and superpotentials involving them are added. This duality transformation was considered implicitly in \refs{\Rains} as a property of integrals which imply equality  of the supersymmetric index of the two dual  models. We will argue that turning on most general superpotentials of the form above, breaking all the flavor symmetry of the gauge models but preserving R symmetry, this duality transformation together with permutations of the eight quarks generates orbits of the Weyl group of $SO(16)$. Then together with the dualities generating $E_7$ orbits the full duality web is that of orbits of the Weyl group of $E_8$.

The note is organized as follows. We start the discussion with a review of dualities transforming $USp(2m)$ models on an $E_7$ orbit. We also make a curious observation about a special property of the $USp(2m)$ with even $m$.  We argue that this model with a particular superpotential flows to a SCFT with $E_7\times U(1)$ symmetry. In the particular case of $USp(4)$ this model sits on the same conformal manifold as the $E_7$ surprise of Dimofte and Gaiotto \DimoftePD.
In
 section three we discuss a duality which relates $USp(2m)$ and $USp(2n)$ models with superpotentials on both sides. In section four we finish by combining the two types of  dualities and explain how they build orbits of the Weyl group of $E_8$. 

\newsec{The $E_7$ orbits of dualities}

The basic theory we consider is a $USp(2n)$ gauge theory with four flavours, $Q_a$ with $a\in\{1,\cdots, 8\}$ in the fundamental representation, and one field in the antisymmetric representation, $X$. 
This model was studied by various authors, in particular see
\refs{\IntriligatorAX,\IntriligatorFF} and references below.
The symmetry is $SU(8)\times U(1 )$.   We call this model ${\frak T}^{(n)}_0$. The charges are summarized in  Table 2.1.

\eqn\chargd{
\vbox{\offinterlineskip\tabskip=0pt
\halign{\strut\vrule#
%%%%%%%%%%%%%%%%%%
&~$#$~\hfil\vrule
&~$#$~\hfil\vrule
&~$#$~\hfil\vrule
&~$#$\hfil\vrule
&~$#$\hfil
&\vrule#
\cr
%%%%%%%%%%%%%%%%%%
\noalign{\hrule}
&  &  USp( 2n)  & SU (8)&    U(1 )  & U(1 )_r &\cr
\noalign{\hrule}
& Q_j              & \; {\bf 2n}    & \; {\bf 8}   & \; -\frac{n-1}4     &   \quad \frac12   &\cr
&  X       & \; {\bf n(2n-1)-1}     & \; {\bf 1}    &\;1  &  \quad  0 &\cr
}
\hrule}} When $n=1$ the field $X$ does not exist and the model is $SU (2 )$ gauge theory with four fundamental flavors. 

The model above has multiple known dual descriptions. The dual description preserving manifestly the most symmetry is given by the same gauge theory but with a collection of singlet fields coupled to gauge invariant operators through the superpotential \CsakiEU,

\eqn\basdu{
W=\sum_{y=1}^n\sum_{i,j} M_{i,j/y}^n q_i q_j \widetilde X^{y-1}\,.
} The charges of the fields are in Table 2.3.

\eqn\chargd{
\vbox{\offinterlineskip\tabskip=0pt
\halign{\strut\vrule#
%%%%%%%%%%%%%%%%%%
&~$#$~\hfil\vrule
&~$#$~\hfil\vrule
&~$#$~\hfil\vrule
&~$#$\hfil\vrule
&~$#$\hfil
&\vrule#
\cr
%%%%%%%%%%%%%%%%%%
\noalign{\hrule}
&  &  USp( 2n)  & SU (8)&    U(1 )  & U(1 )_r &\cr
\noalign{\hrule}
& q_j              & \; {\bf 2n}    & \; \overline{{\bf 8}}   & \; -\frac{n-1}4     &   \quad \frac12   &\cr
&  \widetilde X       & \; {\bf n(2n-1)-1}     & \; {\bf 1}    &\;1  &  \quad  0 &\cr
& M_{i,j/y} &\;{\bf 1} &\;{\bf 28} & \frac{n+1}2-y & \quad 1&\cr
}
\hrule}} The map between the operators is,

\eqn\maodi{
Q_iQ_j X^l \;\to M_{ij/n-l}\,,\qquad\, X^l\to \widetilde X^l\,.
} This is a generalization of Intriligator-Pouliot duality \refs{\IntriligatorNE}
 for $n$ equals one case. We can build many other duality frames which will have less symmetry manifest in a similar manner to the $n=1$ case. 
The number of duals is $72=W(E_7)/W(A_7)$. 
To construct these dualities we split the eight fundamental fields into two groups of four. The matter content is then written  in Table 2.5.

\eqn\chargdhye{
\vbox{\offinterlineskip\tabskip=0pt
\halign{\strut\vrule#
%%%%%%%%%%%%%%%%%%
&~$#$~\hfil\vrule
&~$#$~\hfil\vrule
&~$#$~\hfil\vrule
&~$#$~\hfil\vrule
&~$#$~\hfil\vrule
&~$#$\hfil\vrule
&~$#$\hfil
&\vrule#
\cr
%%%%%%%%%%%%%%%%%%
\noalign{\hrule}
&  &  USp( 2n)  & SU (4)_1&SU (4)_2 &   U(1 )_b &  U(1 )  & U(1 )_r &\cr
\noalign{\hrule}
& Q_{1,...,4}             & \; {\bf 2n}    & \; {\bf 4} &{\bf 1}& \;1   & \; -\frac{n-1}4     &   \quad \frac12   &\cr
& Q_{5,...,8}             & \; {\bf 2n}    & \; {\bf 1}  &{\bf \bar 4}& -1  & \; -\frac{n-1}4     &   \quad \frac12   &\cr
&  X       & \; {\bf n(2n-1)-1}     & \; {\bf 1}  & {\bf 1} & 0  &\;1  &  \quad  0 &\cr
}
\hrule}}
We have thirty five choices to perform such splitting. One set of thirty five duals is  
 then given by the following matter content,

\eqn\chargd{
\vbox{\offinterlineskip\tabskip=0pt
\halign{\strut\vrule#
%%%%%%%%%%%%%%%%%%
&~$#$~\hfil\vrule
&~$#$~\hfil\vrule
&~$#$~\hfil\vrule
&~$#$~\hfil\vrule
&~$#$~\hfil\vrule
&~$#$\hfil\vrule
&~$#$\hfil
&\vrule#
\cr
%%%%%%%%%%%%%%%%%%
\noalign{\hrule}
&  &  USp( 2n)  & SU (4)_1&SU (4)_2 &   U(1 )_b &  U(1 )  & U(1 )_r &\cr
\noalign{\hrule}
& q_{1,...,4}             & \; {\bf 2n}    & \; {\bf \bar 4} &{\bf 1}& \;1   & \; -\frac{n-1}4     &   \quad \frac12   &\cr
& q_{5,...,8}             & \; {\bf 2n}    & \; {\bf 1}  &{\bf  4}& -1  & \; -\frac{n-1}4     &   \quad \frac12   &\cr
&  X       & \; {\bf n(2n-1)-1}     & \; {\bf 1}  & {\bf 1} & 0  &\;1  &  \quad  0 &\cr
&M_{/l} & {\bf 1} &{\bf 4} &{\bf \bar 4} & 0 & \frac{n+1}2-l &\quad 1 &\cr
}\hrule},} with superpotential given here,

\eqn\supsdu{
W=\sum_{l=1}^n\sum_{i,j=1}^4 M_{i,j/l} \tilde X^{l-1}q_iq_{j+4}\,.} The mesons map to singlets and ``baryons'' to ``baryons'' as antisymmetric square of ${\bf \bar 4}$ (${\bf 4}$) is the real representation ${\bf 6}$. This is the analogue of \SeibergPQ\ Seiberg duality. Another thirty five  duals have  the superpotential  given by the following,

\eqn\shudfu{
W=\sum_{l=1}^n\sum_{i,j=1,i\neq j}^4 (\hat M_{i,j/l} q_i q_j \widetilde X^{l-1}+\hat M'_{i,j/l}q_{j+4}q_{i+4} \widetilde X^{l-1})\,,} with the matter content given in table 2.9. 

\eqn\chargd{
\vbox{\offinterlineskip\tabskip=0pt
\halign{\strut\vrule#
%%%%%%%%%%%%%%%%%%
&~$#$~\hfil\vrule
&~$#$~\hfil\vrule
&~$#$~\hfil\vrule
&~$#$~\hfil\vrule
&~$#$~\hfil\vrule
&~$#$\hfil\vrule
&~$#$\hfil
&\vrule#
\cr
%%%%%%%%%%%%%%%%%%
\noalign{\hrule}
&  &  USp( 2n)  & SU (4)_1&SU (4)_2 &   U(1 )_b &  U(1 )  & U(1 )_r &\cr
\noalign{\hrule}
& q_{1,...,4}             & \; {\bf 2n}    & \; {\bf 4} &{\bf 1}& -1   & \; -\frac{n-1}4     &   \quad \frac12   &\cr
& q_{5,...,8}             & \; {\bf 2n}    & \; {\bf 1}  &{\bf  \bar 4}& 1  & \; -\frac{n-1}4     &   \quad \frac12   &\cr
&  X       & \; {\bf n(2n-1)-1}     & \; {\bf 1}  & {\bf 1} & 0  &\;1  &  \quad  0 &\cr
&\hat M'_{/l} & {\bf 1} &{\bf 1} &{\bf 6} & -2 & \frac{n+1}2-l &\quad 1 &\cr
&\hat M_{/l} & {\bf 1} &{\bf 6} &{\bf 1} & 2 & \frac{n+1}2-l &\quad 1 &\cr
}\hrule}} The``baryons'' map to the singlets and mesons to mesons.
This is an analogue of the duality  \CsakiCU\
 discussed by Csaki, Schmaltz, Skiba, and Terning. The last two sets of dual descriptions were considered in \SpiridonovZR.

\

\

It is convenient to encode the dualities as transformations on an ordered set of fugacities for the different symmetries. We will parametrize the symmetries as follows using supersymmetric index nomenclature.
Remember that a chiral field of R charge $r$ contributes to the index as $\Gamma_e(  ( q p)^{\frac{r}2} h)$ \refs{\KinneyEJ,\DolanQI} with $h$ being the fugacity of the $U(1)$ symmetry under which the field transforms. We will then denote by $u_i=(q p)^{\frac{1}{4}} h_i t^{-\frac{n-1}{4}}$ the weight for the $i$th quark. Here the fugacity  $t$ is for the $U(1)$ symmetry, and fugacities $h_j$  for $SU(8)$ (then $\prod_{k = 1}^8 h_k=1$). 
We can take the parameters $t$ and $u_k$ to be general with one constraint coming from anomaly cancelation,
 $t^{2n-2}\prod_j u_j=(q p)^2$. 
We define an ordered set  of fugacities parametrizing the gauge sector of the $USp(2n)$ theory to be $(u_1,u_2,u_3,u_4,u_5,u_6,u_7,u_8,n,t)$.   Denote $u_+^4=\prod_{j=1}^4 u_j$ and $u_-^4=\prod_{j=5}^8 u_j$. Then the three dualities we discussed imply the following transformations of this set,

\eqn\fldu{
\eqalign{& (u_1,u_2,u_3,u_4,u_5,u_6,u_7,u_8,n,t)\to (\frac{u_+u_-}{u_1},\frac{u_+u_-}{u_2},\frac{u_+u_-}{u_3},\frac{u_+u_-}{u_4},\frac{u_+u_-}{u_5},\frac{u_+u_-}{u_6},\frac{u_+u_-}{u_7},\frac{u_+u_-}{u_8}, n, t)\,\cr&
(u_1,u_2,u_3,u_4,u_5,u_6,u_7,u_8,n, t)\to (\frac{u_+^2}{u_1},\frac{u_+^2}{u_2},\frac{u_+^2}{u_3},\frac{u_+^2}{u_4},\frac{u_-^2}{u_5},\frac{u_-^2}{u_6},\frac{u_-^2}{u_7},\frac{u_-^2}{u_8}, n,t)\,\cr
&(u_1,u_2,u_3,u_4,u_5,u_6,u_7,u_8,n,t)\to  ( \frac{u_-}{u_+} u_1,\frac{u_-}{u_+} u_2,\frac{u_-}{u_+} u_3,\frac{u_-}{u_+} u_4,\frac{u_+}{u_-} u_5,\frac{u_+}{u_-} u_6,\frac{u_+}{u_-} u_7,\frac{u_+}{u_-} u_8,n,t)}
} These transformations together with permutations of the first eight terms generate the Weyl group of $E_7$.

\

\noindent It has been observed by Dimofte and Gaiotto \DimoftePD\
 that combining two copies of  $USp(2)$ theories by coupling the gauge invariant operators as $Q_j Q_i q_j q_i$, the theory has a point on the conformal manifold in the IR with, at least, $E_7$ symmetry. This fact was related in  \skm\
 to a statement that the two copies of $USp(2)$ with that superpotential can be obtained by compactification of the E string theory on a torus.
 We will now discuss here a generalizations of the former fact to higher rank.

\

\subsec{$E_7\times U(1 )$ surprise}

Consider ${\frak T}^{(m)}_0$ and assume  $m$ even. Turn on superpotential,
$$ \sum_{j=1}^{m/2} Q_l Q_i X^{j-1} M_{il/j} +\sum_{i=2}^{m} X^ix_i \,.$$
We claim this model is self dual under the dualities we have considered. Note that under all the dualities it is either that $Q_i Q_j X^{l-1}$ maps to  itself or to $M_{ij/n-l}$.  The powers of   $X$ map to the same powers on the dual side. This implies that the superpotential maps to itself under the three dualities. The only effect of the duality is the non trivial identification of symmetries. These imply that for example the protected spectrum is invariant under th Weyl group of $E_7$ and thus forms representations of $E_7$. It is then plausible that on some point on the conformal manifold of this model the group enhances to $U(1)\times E_7$. Let us analyze one example in detail.

Consider the $USp( 4)$ theory with the mesons and $X^2$ flipped.\foot{Flipping, that is introducing chiral fields $\phi_{\frak O}$ and coupling them to a theory as $\phi_{\frak O} {\frak O}$ with ${\frak O}$ being an operator to be removed in the IR, is a standard technique in CFT.  For recent discussion of some aspects of this procedure see \refs{\BenvenutiLLE,\BenvenutiKUD}.
} That is the superpotential is $Q_l Q_m M_{m l}+X^2 x$ with $M$ and $x$ gauge singlet fields.
The superconformal R symmetry derived by a maximization \refs{\IntriligatorJJ} assigns  R charge zero to $X$ and R charge half to the quarks. The superconformal $(a,c)$ anomalies coincide with two copies of $SU (2 )$ theories glued together with a superpotential, the $E_7$ surprise model. This suggests that the $USp(4 )$ model sits on the same conformal manifold. In particular 
as the superconformal R charge of $X$ is zero,\foot{This will cease to be the case for higher rank, and in particular there will be no conformal manifold.}  giving it a vacuum expectation value takes us on the conformal manifold of that model\foot{More specifically, an expectation value for $X$ is forbidden by both the F-term and D-term conditions. To turn it on we must deform the superpotential by a term linear in $x$, in which case the F-term conditions forces an expectation value. This operator has conformal R charge $2$ and so this is a marginal deformation in the SCFT.}. Giving such an expectation value Higgses the $USp( 4)$ gauge group to $SU(2)^2$. This generates the $E_7$ surprise model if the mesons are flipped. We thus expect the $USp(4 )$ model to have $E_7$ symmetry somewhere on 
its conformal manifold. The
 model has $U(1)\times SU(8)$ symmetry visible in the Lagrangian and if we study the index we see that the symmetry is enhanced to $E_7\times U(1)$. The index is given by,

\eqn\indio{
1+{\bf 56} t^{\frac12} (p q)^{\frac12}   +{\bf 56} t^{\frac12}(q p)^{\frac12}(q+p)+({\bf 1463}t+\frac1{t^2}-{\bf 133}-1)q p+\dots\,.
}  In particular we see explicitly the conserved current of $E_7$ appearing at order $q p$ in the expansion of the index (see \refs{\BeemYN}). 
The conserved currents multiplet contributes at order $q p$ as a fermionic operator, $-{\bf 133}$. It is worth verifying what are the operators giving such a contribution. Let us list all the operators which contribute at order $q p$ and have vanishing charge under the $U(1)$ symmetry,

\eqn\symue{\eqalign{
&\bar \psi_i Q_j,\qquad \bar \psi^M_{ji} Q_lQ_m X\, ,\qquad
  \bar \psi^M_{ij} M_{lm}\,
\, , \qquad
Q_j Q_i M_{lm}\,     ,
\qquad
Q_i Q_l Q_j Q_k X\,,\cr &
\lambda\lambda\,,\qquad 
\bar \psi^X X\,,\qquad
x X^2\,   , 
\qquad 
\bar \psi^x x\,.}
} Here $\lambda$ is the gaugino. The fields $\psi_i$ are fermionic partners of $Q_i$ and the fields $\psi^L$ are fermionic partners of fields $L$ with $L$ being one of $(M, X, x)$.
The operator $\bar\psi_i Q_j$ forms the ${\bf 63}+1$ of  $SU(8)\times U(1)$ and to construct the adjoint of $E_7$ we also need the ${\bf 70}$, fourth completely antisymmetric power of the ${\bf 8}$, in addition to the ${\bf 63}$.
The operators $Q_i Q_l M_{k l} $ and $\bar \psi^M_{km} M_{ij}$ cancel each other in the index computation as a consequence of the chiral ring relation. The operator $Q_i Q_j Q_k Q_n X$ is in the representation ${\bf 378}+{\bf 336}$ of $SU(8)$. In particular this lacks the rank $4$ antisymmetric representation of $SU(8)$. This arises as to get it one has to contract the $Q$'s antisymmetrically in the flavor index, but, since these are bosonic fields, they must then be contracted antisymmetrically also in the gauge indices. However the fourth totally antisymmetric product of the ${\bf 4}$ of $USp(4)$ is a singlet, and so the product with $X$ then cannot be made gauge invariant. The operator $\bar\psi^M_{lk}Q_i Q_j X$ is in the ${\bf 28}\times \bf{   28} = {\bf 336}+{\bf 378}+{\bf 70}$. The operator which gives us the required $-{\bf 70}$ is then obtained from $Q_l Q_m X \bar \psi^M_{ik} $. The relevant operators in the ${\bf 56}$ are constructed from $M_{ij}$ which
 form the ${\bf \overline{28}}$ of $SU(8)$ and from $X Q_iQ_j$ which forms the ${\bf 28}$. Both operators have charge $\frac12$ under the $U(1)$. We note that the fact  that at order $ q p$ we see $-{\bf 133}-1$, assuming that the theory flows to interacting SCFT in the IR, is a proof, following from the superconformal representation theory \BeemYN,  that the symmetry of the fixed point enhances to at least $U(1)  \times E_7$.

\

\newsec{Rank changing duality}

We consider a deformation of ${\frak T}^{(n)}_0$ by a superpotential term, 

\eqn\supm{
W^{n}_m= Q_1 Q_2 X^{m}\,.
} The theory then will be denoted by ${\frak T}^{(n)}_m$.\foot{ Note that because of the anomaly condition, the superpotential $Q_3Q_4Q_5Q_6Q_7Q_8X^{2n-2-m}$ has R charge $4$ minus the R charge of $W^{n}_m$ and opposite charges under other symmetries. This implies that at least one of these operators is relevant if $m<2n-2$ and that the two terms are marginal in IR. }
 We claim  that if $m$ is bigger than $n$ ${\frak T}^{(n)}_m$ is dual to ${\frak T}^{(m)}_n$ with the additional superpotential and singlet fields,

\eqn\asup{
\Delta W^{m}_n =\sum_{k=n+1}^{m}x_{k} \widetilde X^{k} +\sum_{k=1}^{{m-n}}\sum_{2<i<j<9}q_i q_j M_{ij/k}\widetilde X^{k-1}\,.
} 
The symmetries of the theories are $SU (6)\times SU (2 )\times U(1 )$ and the R-symmetry. When we consider $n$ to be one, the symmetry should enhance to $SU (8)$ though it is not apparent on the $USp(2m)$ side of the duality. The representations under non-abelian symmetries are the same across the duality.  The charges on the $USp(2n)$ side are detailed in Table 3.3.

\eqn\charg{
\vbox{\offinterlineskip\tabskip=0pt
\halign{\strut\vrule#
%%%%%%%%%%%%%%%%%%
&~$#$~\hfil\vrule
&~$#$~\hfil\vrule
&~$#$~\hfil\vrule
&~$#$~\hfil\vrule
&~$#$\hfil\vrule
&~$#$\hfil
&\vrule#
\cr
%%%%%%%%%%%%%%%%%%
\noalign{\hrule}
&  &  USp(2n)  & SU (6)&   SU (2 )  & U(1 )  & U(1 )_r &\cr
\noalign{\hrule}
%%%%%%%%%%%%%%%%%%
&  (Q_{1},Q_2)        & \; {\bf 2n}    & \; {\bf 1}    &\; {\bf  2} & \quad \;-\frac12m   &  \quad  1   &\cr
& Q_{3,...,8}              & \; {\bf 2n}    & \; {\bf 6}   & \; {\bf 1}   & \quad -\frac1{6}(2n-m-2)    &   \quad \frac13   &\cr
&  X       & \; {\bf n(2n-1)-1}     & \; {\bf 1}    &\; {\bf  1} & \qquad\;\,\;1 &  \quad  0 &\cr
}
\hrule}}
On the $USp(2m) $ side  we obtain the charges in Table 3.4.

\eqn\charg{
\vbox{\offinterlineskip\tabskip=0pt
\halign{\strut\vrule#
%%%%%%%%%%%%%%%%%%
&~$#$~\hfil\vrule
&~$#$~\hfil\vrule
&~$#$~\hfil\vrule
&~$#$~\hfil\vrule
&~$#$\hfil\vrule
&~$#$\hfil
&\vrule#
\cr
%%%%%%%%%%%%%%%%%%
\noalign{\hrule}
&  &  USp(2m)  & SU (6)&   SU (2 )  & U(1 )  & U(1 )_r &\cr
\noalign{\hrule}
%%%%%%%%%%%%%%%%%%
&  (q_{1},q_2)        & \; {\bf 2m}    & \; {\bf 1}    &\; {\bf  2} & \quad \;-\frac12n   &  \quad  1   &\cr
& q_{3,...,8}              & \; {\bf 2m}    & \; {\bf 6}   & \; {\bf 1}   & \quad -\frac1{6}(2m-n-2)    &   \quad \frac13   &\cr
&  \widetilde X       & \; {\bf m(2m-1)-1}     & \; {\bf 1}    &\; {\bf  1} & \qquad\;\,\;1 &  \quad  0 &\cr
&  M_{/y}       & \; {\bf 1}     & \; {\bf 15}    &\; {\bf  1} & \qquad\;\,\;-\frac13(n-2m+3y-1) &  \quad  \frac43&\cr
&  x_y       & \; {\bf 1}     & \; {\bf 1}    &\; {\bf  1} & \qquad\;\,\;-y &  \quad  2 &\cr
}
\hrule}} The map of the operators is as follows, 

\eqn\mgfyure{\eqalign{
&X^j\to \widetilde X^j\,, \qquad \;\;\, i, \, l\neq 1,2 \quad Q_iQ_l X^{j} \to q_i q_l \widetilde X^{m-n+j}\,,\cr&
Q_{1,2}Q_{l>2} X^{j-1} \to q_{1,2} q_{l>2} \widetilde X^{j-1}\,,\qquad\, \cr &
j\leq m-n\;\, Q_1Q_2 X^{j-1} \to x_{m-j+1}\,,\qquad \,\,   \, j>m-n \;\, Q_1Q_2 X^{j-1} \to q_1 q_2 \widetilde X^{j-1-m+n}\,,\cr&
j\leq n-1\;\;\;\,\,\, (Q^4)_{ck} X^{j-1} \to M_{ck/n-j}\,.}}
The supersymmetric index of the two sides of the duality agrees as was shown by Rains \Rains. The fact that the index agrees guarantees in particular that the anomalies agree and that the  protected  operators map to each other. Nevertheles, let us detail the anomalies here.
 We can encode anomalies involving abelian symmetries in the trial $c$ and $a$ anomalies. Defining  $R=R'+s q$ with $R'$ and $q$ the R symmetry and the $U(1 )$ charge in the tables above, with $s$ a parameter, the conformal anomalies are,

\eqn\cogti{\eqalign{& a(s) = \frac{1}{32} \left(s^3 \left(-3 \left(m^2+4 m-2\right) n^2+6 (m+2) n^3-(m-1)^2 (4 m+5) n-4 n^4-9\right)\right.-\cr& \;\,3 \left.s^2 \left(\left(2 m^2+8 m-1\right) n+(2-8 m) n^2+8 n^3-9\right)+12 s \left(2 m n+n^2+n-2\right)+4 n+6\right)\cr & c(s) =  \frac{1}{32} \left(s^3 \left(-3 \left(m^2+4 m-2\right) n^2+6 (m+2) n^3-(m-1)^2 (4 m+5) n-4 n^4-9\right)\right.-\cr&\;3\left. s^2 \left(\left(2 m^2+8 m-1\right) n+(2-8 m) n^2+8 n^3-9\right)+2 s\left(3 (4 m+1) n+8 n^2-11\right)+16 n+4\right)\,.}}
A more symmetric way to think about the duality is to define the model ${\frak T}^{(n)}_0$ with the superpotential given by the following. 

\eqn\asup{
W^{(n)}_0 =\sum_{k=2}^{n}x_{k} \widetilde X^{k} +\sum_{k=1}^{n}\sum_{2<i<j<9}q_i q_j M_{ij/k}\widetilde X^{k-1}\,.
} We denote this theory as $\bar{\frak I}^{(n)}_0$ and the model with $Q_1Q_2 X^m$ as $\bar {\frak I}^{(n)}_m$. We then claim that $\bar{\frak I}^{(n)}_m$ is dual to $\bar{\frak I}^{(m)}_n$.
 
Let us define the index of a theory $\bar{\frak I}^{(n)}_m$ to be  $I^m_n(u_1,u_2,\cdots,u_8,t)$ with the condition coming from anomalies $t^{2n-2} \prod_{i=1}^8 u_i=p^2 q^2$.  Here again $u_i$ are the weights of the quarks as defined in the previous section and $t$ is the fugacity for the $U(1)$ under which the antisymmetric chiral field is charged. Turning on the superpotential $Q_1Q_2X^m$ we identify $u_1 u_2 t^m= p q$. If we define $u^2= p q\frac{t^{-n}}{u_2u_1}$,  the duality implies that the index satisfies the following identity,

\eqn\duiriou{
I^m_n(u_1,u_2,u_3,u_4,u_5,u_6,u_7,u_8,t) = I_m^n(u_1 u,u_2 u, \frac{u_3}u,\frac{u_4}{u},\frac{u_5}{u},\frac{u_6}{u},\frac{u_7}{u},\frac{u_8}u,t)\,.
} This can be derived from the map of symmetries between the two dualities.

\

\subsec{Duals of $SU (2 )$ SQCD with four flavors}

Consider an example of the rank changing duality with $n=1$ and general $m$. The theory on one side  is always $USp( 2)=SU (2 )$ SQCD with eight fundamental chiral fields and no superpotential. On the dual side we have a $USp( 2m)$ model with  superpotential $W=q_2q_1 \widetilde X$ and other terms involving singlet fields we have discussed. 
We then have an infinite number of duals for the $SU(2)$ theory with eight fundamental chiral fields.  The fact that this model has an infinite number of duals
 is not surprising \AgarwalVLA, but the surprising point is that the duals are rather simple.

Let us work out a simple example. We consider $m$ to be two. At the fixed point of the gauge theory with no superpotential the operator $q_2q_1  \widetilde X$ has R charge $1.15749$ and thus is relevant. The operator $\widetilde X^2$ violates the unitarity bound and needs to be decoupled by introducing a flip $x$ appearing in the superpotential as $x \widetilde X^2$. After decoupling the $\widetilde X^2$ operator we get that the R charge of $\widetilde X q_1 q_2$ is $1.15331$. We turn it on and flow to a new fixed point. At that point the operators ($ i,j\neq 1,2$) $q_i  q_j$ violate the unitarity bound and need to be decoupled by introducing flippers. After this there are no operators violating unitarity bounds and we get precisely the superpotential we obtain from the duality. Thus $USp(4)$ theory with the superpotenial flipping $\widetilde X^2$ and $q_i q_j$ ($i,j\neq 1,2$), and superpotential term 
$q_2 q_1\widetilde X$ flows to $USp(2)$ with no superpotential.  For low values of $m$ we can repeat such an analysis though for higher values it becomes rather intricate.

\

\newsec{The duality orbit}

We consider the more general superpotential for a $USp( 2{\frak  a}_9  )$ theory, 

\eqn\gsu{
W_{{\frak a}_9;{\frak a}_1,\cdots, {\frak a}_8}=\sum_{i\neq j} Q_i Q_j X^{{\frak a}_i+{\frak a}_j}\,.} This superpotential breaks all the flavor symmetry but preserves the R symmetry\foot{Here we assume that ${\frak a}_9 \neq 1+\sum_{j=1}^8 \frac{{\frak a}_j}{2}$. If this is not true then there is no R symmetry, but instead there is an anomaly free $U(1)$ global symmetry. We shall not discuss this case in any detail.}. The parameters ${\frak a_j}$ are either all integer or all half integer.   The R charges are,

\eqn\chargai{
r_{Q_i}= 1-{\frak a}_i r_X\,,\qquad\, r_X=\frac4{2-2{\frak a}_9+\sum_{j=1}^8 {\frak a}_j}.
}  Some operators violate the unitarity bounds for general choices of ${\frak a}_i$ and  need to be decoupled.  
We define the parameters $u_i$ and $t$ to be as in previous sections,

\eqn\hyuioie{
 u_i=(p q)^{\frac12}t^{-{\frak a}_i}\,.
} Note also that now $t=(q p)^{\frac12 r_X}$.
 The index is given by

\eqn\ertyue{
I^{{\frak a}_1+{\frak a}_2}_{{\frak a}_9}((p q)^{\frac12}t^{-{\frak a}_1},\cdots,(p q)^{\frac12}t^{-{\frak a}_8},t)\,.
} The duality of the previous section then implies that this index is equal to,

\eqn\duim{
I^{{\frak a}_9}_{{\frak a}_1+{\frak a}_2}((p q)^{\frac12}t^{-\frac12({\frak a}_1-{\frak a}_2+{\frak a}_9)},(p q)^{\frac12}t^{-\frac12({\frak a}_2-{\frak a}_1+{\frak a}_9)}, (p q)^{\frac12}t^{-\frac12(2{\frak a}_3+{\frak a}_2+a_1-{\frak a}_9)},\cdots,t)}
We parametrize again the theory by the nine numbers, which here are integers (or half integers),
 $({\frak a}_9,{\frak a}_1,{\frak a}_2,{\frak a}_3,{\frak a}_4,{\frak a}_5,{\frak a}_6,{\frak a}_7,{\frak a}_8)$. The duality transforms 

\eqn\duatryu{
({\frak a}_9,{\frak a}_1,{\frak a}_2,{\frak a}_3,{\frak a}_4,{\frak a}_5,{\frak a}_6,{\frak a}_7,{\frak a}_8)\to
({\frak a}_1+{\frak a}_2, \frac12({\frak a}_1-{\frak a}_2+{\frak a}_9),\frac12({\frak a}_2-{\frak a}_1+{\frak a}_9),\frac12(2{\frak a}_3+{\frak a}_1+{\frak a}_2-{\frak a}_9),\cdots)\,   .}
The transformations with permutations of the last eight numbers generate the Weyl group of $SO(  16)$. 

We can also act with the duality generating the $E_7$ orbit. 
Denote  ${\frak a}_+=\frac12\sum_{i=1}^{4}{\frak a}_i$, ${\frak a}_-=\frac12\sum_{i=5}^8{\frak a}_i$, ${\frak a}=\frac12({\frak a}_-+{\frak a}_+)$, ${\frak a}'=\frac12({\frak a}_--{\frak a}_+)$. Then the three dualities generating the $E_7$ orbit  imply the following transformations of this set,

\eqn\Jtra{\eqalign{
&({\frak a}_9,{\frak a}_1,{\frak a}_2,{\frak a}_3,{\frak a}_4,{\frak a}_5,{\frak a}_6,{\frak a}_7,{\frak a}_8)\to\cr&\,
({\frak a}_9,{\frak a}_+-{\frak a}_1,{\frak a}_+-{\frak a}_2,{\frak a}_+-{\frak a}_3,{\frak a}_+-{\frak a}_4,{\frak a}_--{\frak a}_5,{\frak a}_--{\frak a}_6,{\frak a}_--{\frak a}_7,{\frak a}_--{\frak a}_8)\to\cr&
\,({\frak a}_9,{\frak a}-{\frak a}_1,{\frak a}-{\frak a}_2,{\frak a}-{\frak a}_3,{\frak a}-{\frak a}_4,{\frak a}-{\frak a}_5,{\frak a}-{\frak a}_6,{\frak a}-{\frak a}_7,{\frak a}-{\frak a}_8)\to\cr& ({\frak a}_9,{\frak a}'+{\frak a}_1,{\frak a}'+{\frak a}_2,{\frak a}'+{\frak a}_3,{\frak a}'+{\frak a}_4,-{\frak a}'+{\frak a}_5,-{\frak a}'+{\frak a}_6,-{\frak a}'+{\frak a}_7,-{\frak a}'+{\frak a}_8)\,.}} These transformations and permutations of  last eight elements generate the action of the Weyl group of $E_7$ as was observed previously. The $SO(16)$ Weyl group with the $E_7$ transformations  gives the Weyl group of $E_8$.
Note  that the combination $2-2{\frak a}_9+\sum_{i=1}^8 {\frak a}_i$ is invariant under all the transformations.

The transformations can generate a full orbit of the Weyl group of $E_8$ for general values of the parameters. For special values they might not act faithfully and generate smaller orbits. For example take all ${\frak a}_i $ for $i=1...8$ to be $\frac12{\frak a}_9$. Then the $SO(16)$ transformations become self dualities.

There is an issue of whether or not the theories defined via the superpotential \gsu\
 are indeed all unique and define a non-trivial SCFT. The basic problem is that the superpotentials seem irrelevant in the IR so one may fear that the RG flow they generate is trivial. A related issue is that for generic values of ${\frak a}_i $ there will be operators with R charge below the unitarity bound. To deal with the latter problem we can add, to all sides of the duality, flipping fields that remove these operators. The exact number of operators required may depend on the value of $2-2{\frak a}_9+\sum_{i=1}^8 {\frak a}_i$, but one can still perform this action so as to result in an orbit where all operators are above the unitarity bound. Here it is important that the R symmetry is fixed so this does not lead to any flow that may cause more operators to go below the unitarity bound. Therefore this statement can be phrased as an identity between a collection of theories
  where all operators are above the unitarity bound, and so there is no contradiction with them flowing to an SCFT. Of course we cannot rule out the possibility that in some cases the flow may be trivial and the superpotential are truly irrelevant rather then dangerously irrelevant.  
  
It will be interesting to understand whether the fact that there are theories residing on an orbit of the Weyl group of $E_8$ implies that there is a model with $E_8$ symmetry, and in which particular way this fact is related to compactifications of six dimensional $(1,0)$ models.

\

\

\noindent{ {\bf {Acknowledgments}}

We would like to thank O. Aharony, D. Gaiotto, K. Intriligator, H. C.~Kim, Z.~Komargodski,   and C. Vafa for useful comments and discussions.
GZ is supported in part by  World Premier International Research Center Initiative (WPI), MEXT, Japan.  SSR is  a Jacques Lewiner Career Advancement Chair fellow. The research of SSR was also supported by Israel Science Foundation under grant no. 1696/15 and by I-CORE  Program of the Planning and Budgeting Committee.

\listrefs

\end